\documentclass[12pt,preprint]{aastex}
\usepackage{amssymb, amsmath}
\usepackage{color}
\usepackage{morefloats}
\usepackage{hyperref}

\shortauthors{Neugent et al.}

\begin{document}

\title{Binary Red Supergiants: \\A New Method for Detecting B-type Companions}

\author{Kathryn F.\ Neugent\altaffilmark{1,2}, Emily M.\ Levesque\altaffilmark{1}, and Philip Massey\altaffilmark{2,3}}

\altaffiltext{1}{Department of Astronomy, University of Washington, Seattle, WA, 98195 ; kneugent@uw.edu; emsque@uw.edu}
\altaffiltext{2}{Lowell Observatory, 1400 W Mars Hill Road, Flagstaff, AZ 86001; phil.massey@lowell.edu}
\altaffiltext{3}{Department of Physics and Astronomy, Northern Arizona University, Flagstaff, AZ, 86011-6010}

\begin{abstract}
With the exception of a few well-known and studied systems, the binary population of red supergiants (RSGs) remains relatively uncharacterized. Famous systems such as VV Cep, 31 Cyg and $\zeta$ Aur contain RSG + B star binaries and here we explore whether B stars are the main type of companion we expect from an evolutionary point of view. Using the Geneva evolutionary models we find that this is indeed the case. However, few such systems are known, and we use model spectra to determine how easy such binaries would be to detect observationally. We find that it should be quite difficult to hide a B-type companion given a reasonable signal-to-noise in the optical / blue portion of the spectrum. We next examine spectra of Magellanic Cloud RSGs and newly acquired spectra of Galactic RSGs looking for new systems and refining our conclusions about what types of stars could be hidden in the spectra. Finally, we develop a set of photometric criteria that can help select likely binaries in the future without the overhead of large periodic or spectroscopic surveys.
\end{abstract}

\section{Introduction}
It has long been recognized that a significant fraction of stars are found in binaries.  Studies of un-evolved massive stars typically find an observed binary fraction of 30-35\% for O-type stars in relatively short period systems \citep{Garmany80, Sana30Dor}. When long-period systems are included, this percentage rises and approaches 70\% or more \citep{Gies08, SanaSci}. However, it is not clear to what extent these long-period systems are important from an evolutionary perspective (see, for instance, discussion in \citealt{MasseyEngland}).  One key to answering this is studying evolved massive stars. Research has placed constraints on the binary fraction of evolved massive stars, the Wolf-Rayet stars, at around 30\% \citep{WRbins}. However, less is known about the intermediate stars such as the red supergiants (RSGs). The binary fraction of such stars has not been well studied and thus very few binary RSGs are known (the most famous being VV Cephei, $\zeta$ Auriga, and 31 Cygni). It is not well understood whether the rarity of known binary RSGs is a consequence of a low binary fraction or simply due to some limitation in the observations.

Binary stars can be detected using several different methods; perhaps the simplest and most widely-used approach is to search for ellipsoidal variations in a star's light curve. However, detecting binary RSGs in this manner has proven to be quite difficult. RSGs themselves have quite large photometric variations, often on the order of 1 magnitude or higher. These variations can create their own periodic light curves, often making it difficult to distinguish between an orbiting star or something intrinsic to the RSG itself. For example, RSGs are known to pulsate on the order of 80 to 3500 days \citep{GuoLi}, and large convective cells ($\sim$0.5-2 AU across) and hot spots in their chromospheres have timescales on the order of weeks to months or even years \citep{Chiavassa11, Baron14, Stothers10}. Photometric variability in these stars can also be random due to shocks in the atmosphere which excite emission features and large mass-loss events which generally take place at the end of a RSG's lifetime. All of these variations create difficulties in searching for binary RSGs based on their periodic light curve signatures.

Spectroscopically it is often possible to detect these binaries. Examples include 32 Cyg, a K5Iab + B7V \citep[e.g.][]{McLaughlin1950, Chun1992}, 31 Cyg, a K4Iab + B4IV-V \citep[e.g.][]{Stencel1984, Hansen1944, Bauer2014, Samus09}, $\zeta$ Aur, a K5II + B7V \citep[e.g.][]{Harper2016, Wright1970}, $\beta$ Per, a K0IV + B8V \citep[e.g.][]{Sarna1993}, XX Per, a M4Ib + B7V \citep[e.g.][]{Stothers1971}, VV Cep, a M2Iab + B0-2V \citep[e.g.][]{Cowley1969, Pollmann2017} and others where (in most cases) it is possible to see both the characteristic titanium oxide (TiO) bands of a RSG as well as the upper Balmer lines present within a hot un-evolved B-type star. However, searching for binary RSGs blindly and obtaining sufficient signal to noise (S/N) in the fainter bluer wavelengths to hunt for Balmer lines from potential companions is quite a task, particularly as some of these systems are eclipsing binaries where the B star spectrum is not always visible (e.g. VV Cep, $\zeta$ Aur). Another spectroscopic method, searching using radial velocity variations, comes with its own host of problems as it requires a targeted spectroscopic survey with repeated observations over a prolonged period of time (especially considering periods of binary RSGs are on the order of thousands of days or close to the limit under which interactions occur and disrupt the stars' evolutions). 

In this work we avoid the difficulties of periodic variations and large spectroscopic surveys by developing single-epoch photometric methods for identifing RSG binary candidates. In Section 2 we discuss the types of stars that can exist in a binary system with RSGs. In Section 3 we describe our use of synthetic models to create RSG binary spectra. We then discuss previous observations of RSG binary systems as well as our own efforts to detect such stars in Section 4. In Section 5 we outline the photometric cutoffs necessary to create a sample of RSG binary candidates using colors alone. Finally we conclude and discuss future work in Section 6.

\section{B-Type Stars}
Of the known RSG binary systems (see, for example, the six listed in the Introduction), the vast majority of them contain a B star companion. Could this simply be a byproduct of observational biases? For example, would a bright O star simply dominate the RSG spectrum? Or would an A dwarf spectrum not be visible at all in a binary system? To answer this question, we first examined this from an observational point of view by combining spectra to ``create" different binary systems. We then investigated this issue from an evolutionary point of view by determining what types of stars at different masses would be alive during a RSG's lifetime.

\subsection{First Look: The Atlas}
We first created our own set of spectra showing RSGs in systems with various types of stars. To do this we used a library of stellar spectra created by \citet{JHCAtlas}. This spectral database contains fluxed spectra spanning all types and luminosity classes of Galactic stars. Using this database we were able to make a grid of RSG binary spectra combined with the spectra of different types of stars. While we realize that many of these combinations are unphysical from an evolutionary point of view (as discussed in the next section), we first wanted to investigate this from an observational point of view.

To combine the observed spectra we first needed to make several adjustments. We corrected for the reddening of each spectrum using the CCM law \cite{CCM} where we obtained $V$ using SIMBAD \citep{SIMBAD} and were provided E(B-V) \citep{JHCAtlas}. We then corrected for the distance to each star using the absolute visual magnitudes based on spectral type from \citet{AAQ} and \citet{Humphreys84}. Multiplying the derived distance correction by the spectral flux produced a reddening and distance corrected spectrum. At this point we could simply add the spectra together to get an ``observed" combined spectrum.

\begin{figure}
\epsscale{0.45}
\plotone{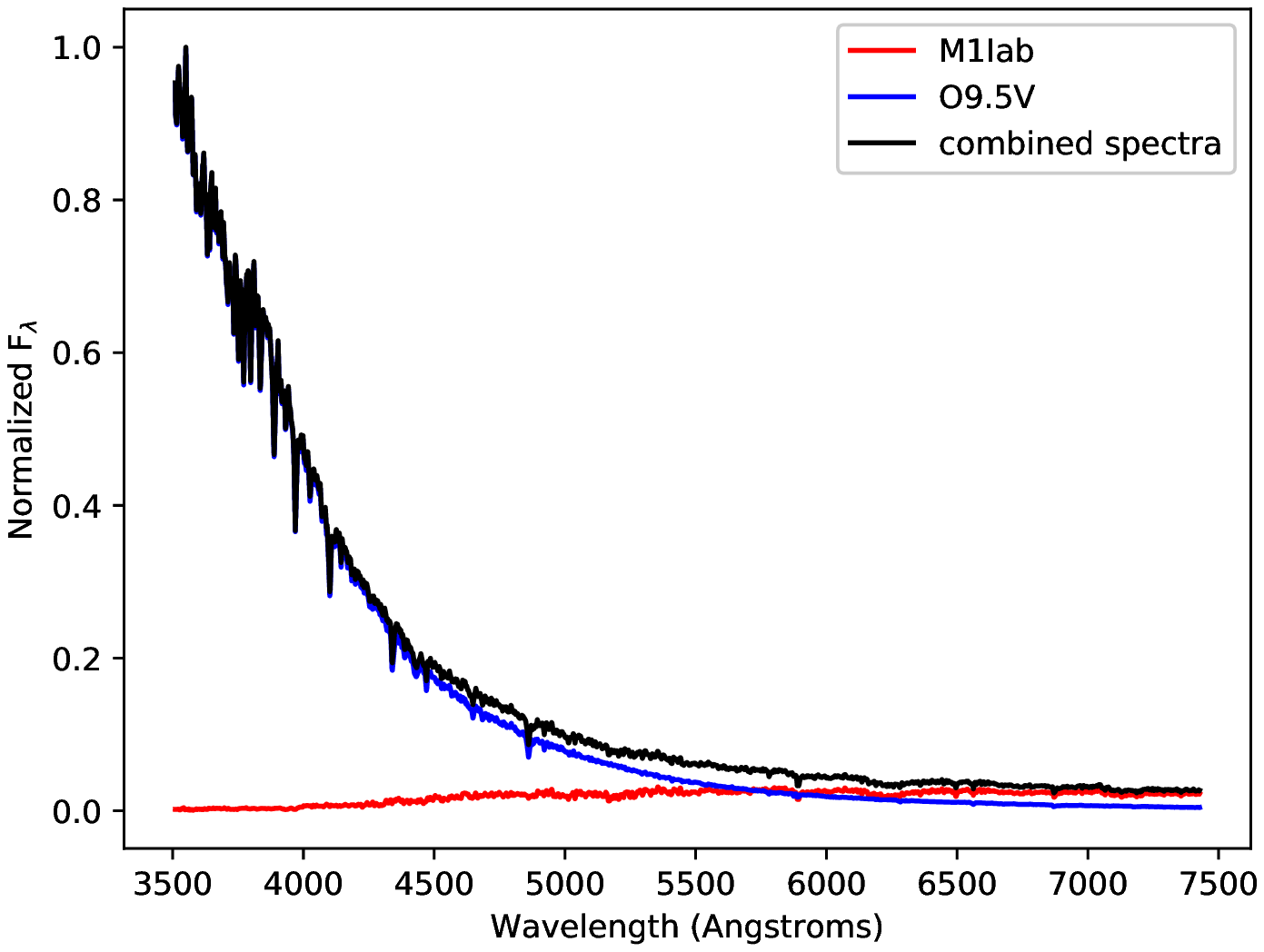}
\plotone{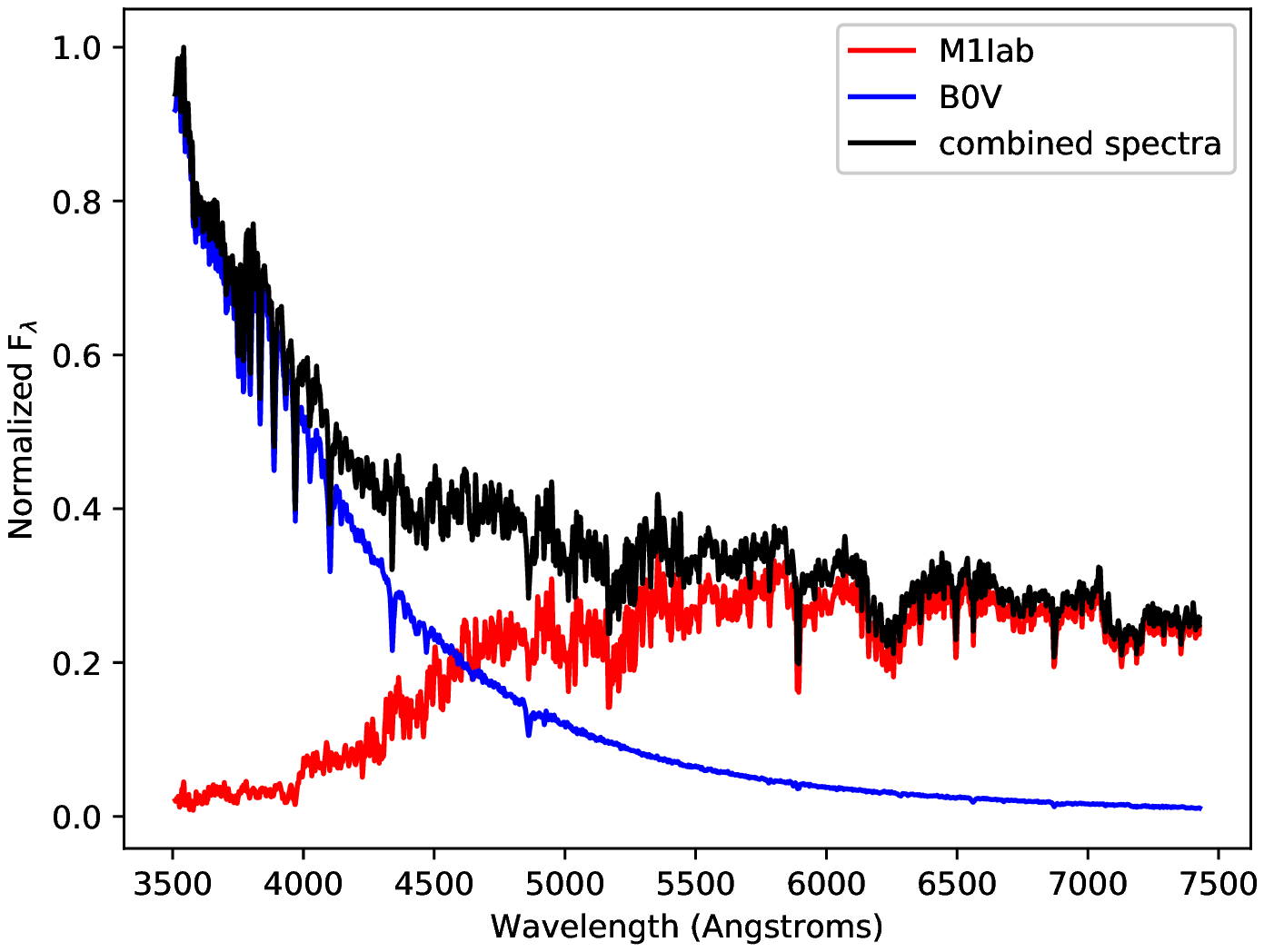}
\plotone{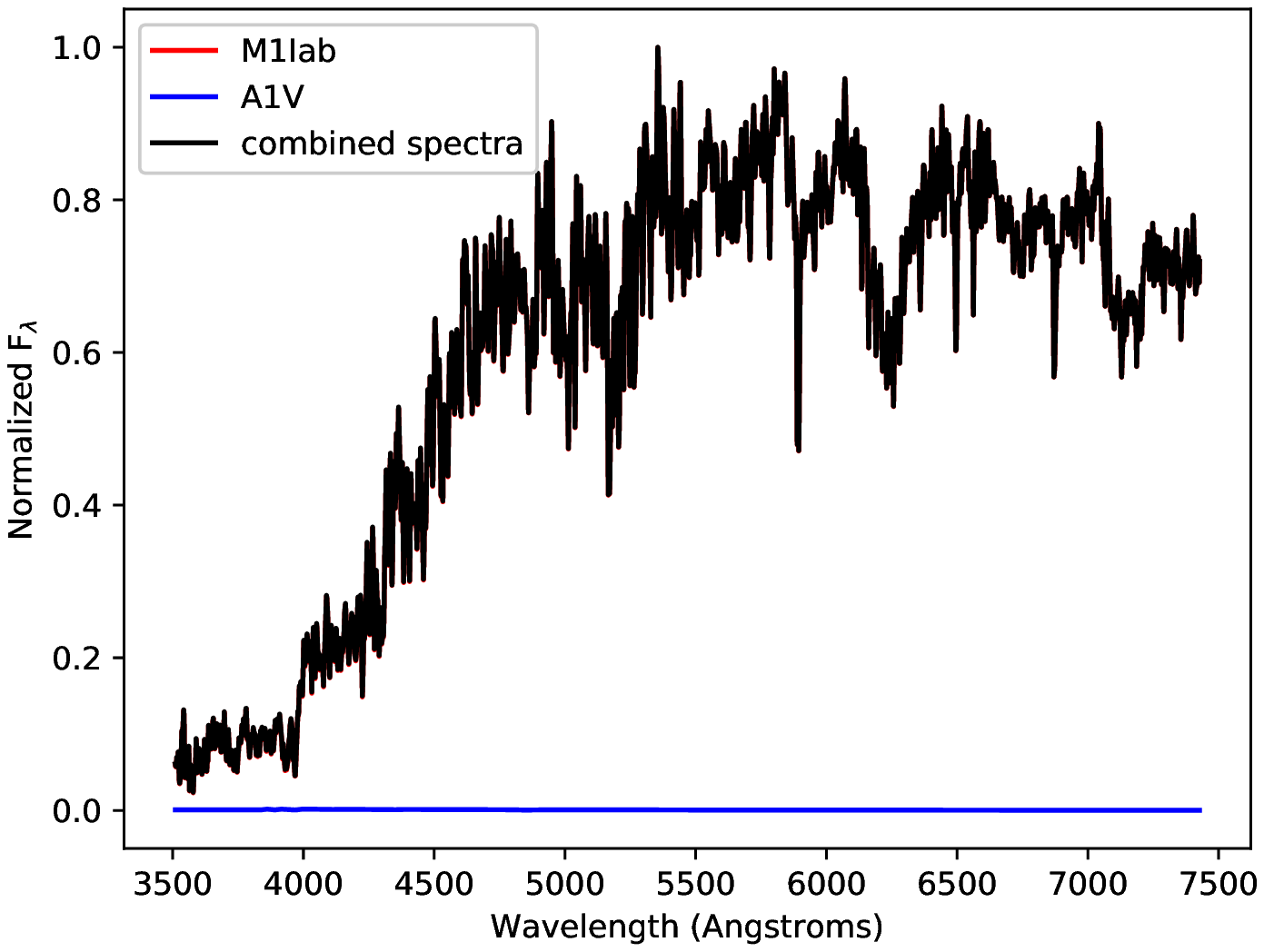}
\plotone{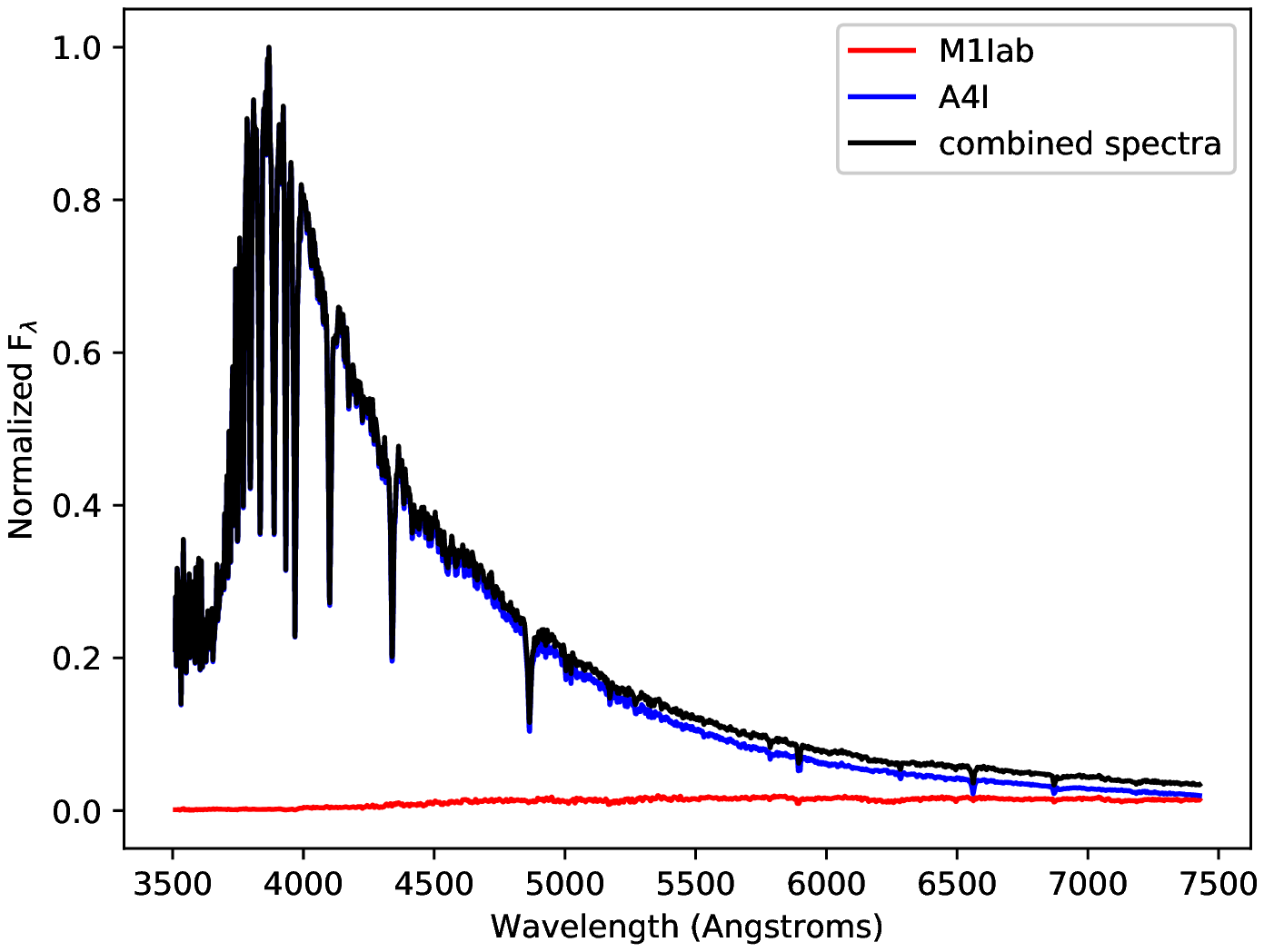}
\caption{\label{fig:JHCspec}A spectrum of V* FZ Per, a M1Iab, combined with the spectra of O9.5V, B0V, A1V and A4I stars. The upper left figure shows the M1Iab spectrum combined with the spectrum of BD+62 0249, an O9.5V. In this figure the O9.5V's spectrum dominates the combined spectrum. The upper right figure shows the M1Iab spectrum combined with the spectrum of HD 3237007, a B0V. In this figure both the blue upper Balmer lines and red TiO lines are clearly visible in the combined spectrum. The bottom left figure shows the M1Iab spectrum combined with the spectrum of HD 116608, an A1V. In this case the M1Iab's spectrum is the only one visible in the combined spectrum. The bottom right figure shows the opposite situation where the A4I spectrum (SAO 12096) dominates over the M1Iab RSG's spectrum. All stars and spectra taken from \citet{JHCAtlas}.}
\end{figure}

Figure~\ref{fig:JHCspec} shows a few examples of spectra combined with the spectrum of a M1Iab, V* FZ Per. \citet{Humphreys84} estimate an $M_V$ of -6.0 for a typical M1Iab (later we redo this investigation for a more complete set of $M_V$s)\footnote{Now that DR2 {\it Gaia} distances are available, better calibrations for the absolute magnitudes of RSGs will doubtless become available, but this is beyond the scope, or needs, of the present project (also, some of the individual stars we use from \cite{JHCAtlas} are actually too bright for reliable {\it Gaia} parallaxes.) The absolute magnitudes we have adopted are consistent with what is known for RSGs where the distances are well known, such as the LMC \citep{Levesque06} and M31/M33 \citep{DroutM33RSGs, MasseyM31RSGs}. We note that for V* FZ Per the agreement is quite good between our approximate method ($M_V = -6$) and the more exact Mv given by the distance Gaia provides ($M_V = -5.6$).}. Note that, as hypothesized above, the O star spectrum (even as a dwarf) simply overpowers the RSG spectrum. In this case, while the SED shows a slight turn-up in the red, this would hardly be noticeable even with perfect flux calibration. Conversely, both the blue B star's spectrum and the red RSG's spectrum is clearly visible in the combined spectrum for the M1Iab+B0V. This combined spectrum's SED is nearly flat, an attribute we rely on later when searching for these stars photometrically. The remaining two figures show the RSG spectrum combined with a spectrum of an A dwarf and an A supergiant. In the A dwarf's case, the RSG spectrum dominates while in the A supergiant's case, the A star dominates. It would be impossible to determine binarity from spectroscopy. Overall, there appears to be a sweet spot when detecting a binary system spectroscopically because in most cases one of the spectra will ``drown out" the other.

\subsection{Evolutionary Models}
While the above exercise is helpful, these stars must co-exist from an evolutionary point of view if they are to be in a binary system with one another. To investigate this we used the Geneva evolutionary models \citep{Geneva, Ekstrom2012}\footnote{Since we are assuming these are non-interacting, wide-period binaries, we believe using single star models is appropriate \citep{CyrilNew}.} to determine the type of star that would exist in a system with a RSG at different mass ranges. We used the evolutionary models to create a grid of models between 0.9 and 29 solar masses. 

We did not go above 29$M_{\odot}$ because we determined by interpolating between the published Geneva tracks\footnote{\url{https://www.unige.ch/sciences/astro/evolution/en/database/syclist/}} that the most massive star that will become a RSG is 29$M_{\odot}$ \cite{Ekstrom2012}. Since the more massive star is going to evolve first, we were justified in starting with 29$M_{\odot}$. We then began with the highest mass model and combined it with the lower mass models and did this iteratively for each mass, decreasing in mass after each run. For example, we first combined the 29$M_{\odot}$ model with the 27, 25, 23, 20, 15, 12, 9, 7, 5, 4, 3, 2, 1, and 0.9$M_{\odot}$ models before then combining the 27$M_{\odot}$ model with the 0.9-25$M_{\odot}$ models and so on. For each of these model combinations, we determined the age at which the higher mass model - the primary in a binary system - would be a RSG (as defined by having $T_{\rm{eff}}$ less than 4500 K) and then determined what type of star the lower mass model would be during the timeframe that the primary would spend as a RSG.

For every one of these combinations we found that the RSG will always be in a system with a B star companion (either a V, III, or I) -- this is the primary evolutionary pairing. To make sense of this, we make the assumption that both of the stars within the system are non-interacting (most likely a wide-period binary) and were formed at the same time. While there are certainly cases where this is not true (for example, a star that is born as a single star and later gravitationally captured by a companion), in general we can assume that both stars in the system formed at the same time. 

\section{Stellar Atmosphere Models}
Knowing that binary RSGs primarily exist in systems with B stars, we next used synthetic spectra from model atmospheres to create a database of B star + RSG synthetic spectra. To do this we relied most heavily on two atmospheric modeling database: BSTAR06 \citep{BSTARS06} and the MARCS models \citep{MARCS}\footnote{We chose the MARCS models over the PHOENIX models \citep{PHOENIX} because they were successfully used by \citet{Levesque05} to redefine the effective temperature scale of RSGs and show that their location on the Hertzsprung-Russell diagram was consistent with evolutionary theory.}.

BSTARS06 is a grid of non-LTE, line-blanketed, plane-parallel, hydrostatic model atmospheres. Temperatures range from 15000 K to 30000 K and include surface gravities ($\log g$) between 1.75 and 4.75 dex. Metallicities range from one tenth to two times solar. There are additional models computed specifically for B supergiants. More information can be found in \citet{BSTARS06}. 

The MARCS models are a grid of one-dimensional, hydrostatic, plane-parallel and spherical LTE model atmospheres. They encompass spectral types F, G, and K. There are 280 specialized spherical RSG models that were used specifically for this work. Their effective temperatures range from 3400 to 4250 K with [cgs] surface gravities between -0.5 and 1 dex. Metallicities range from one-tenth to 3 times solar ([Fe/H]=-1.0 to +0.3) and the micro-turbulence parameter is 5 km s$^{-1}$. More information can be found in \citet{MARCS}.

\subsection{Combination Process}
Over all, 24,640 spectra were created from these models, combining RSGs and B stars of various temperatures and luminosities. These contained B-type stars ranging from 15000 K to 30000 K (in 1000K steps) with luminosity classes of V, III and I; RSGs from 3400 K to 4250 K (in 1000K steps with the exception of the last model) with masses of 10, 15, 20 and 25 $M_{\odot}$; and all of these combinations at reddenings of $A_V$ = 0 to 2.6 (in steps of 0.2). 

We used the solar metallicity models for all stars (though we later investigated the effect of different metallicities on our results). Surface gravities for the BSTAR06 models were varied based on the type and luminosity class of the star (Supergiants ranged from $\log g = 2.50$ dex at the hot end and 3.00 dex at the cooler temperatures. Giants ranged between 3.25 dex to 3.50 dex and dwarfs were set to 4.0 dex) while surface gravities of the MARCS RSG models were set to 0.00 dex. All lines were rotationally broadened. The MARCS models were already rotationally broadened before being distributed while the BSTAR06 models were broadened using 150 km s$^{-1}$ following \citet{Huang06}. 

Before adding the spectra together, we first put their fluxes on the same absolute scale. The surface fluxes for both sets of models are in units of [erg s$^{-1}$ cm$^{-2}$ Angstrom$^{-1}$]. However, before combination we needed to account for their differing luminosities and radii. For the MARCS models, radii and luminosities were taken from the MARCS inputs as given in the model headers. For the B stars we determined the $M_V$ and bolometric corrections based on \citet{Humphreys84} and selected an appropriate surface gravity. We then determined the bolometric luminosities by adopting the absolute visual magnitudes. The radii of each B star model was then calculated using the appropriate luminosity and temperature. Finally, we multiplied the model flux by the appropriate radii. 

Our treatment of dust is relatively simplistic. The spectra were reddened for values ranging between $A_V = 0$ and 2.6 in steps of 0.2 using the CCM law \citep{CCM} as RSGs are known to have $A_V$ values that vary over this range (see, for example, Table 4 of \citealt{Levesque05}). We made the conservative assumption that the B star would be enshrouded in the RSG's dust, although this would clearly depend upon the orbital separation.  If the B star were less reddened than the RSG then its spectrum in the blue would be even more pronounced.

\begin{figure}
\epsscale{0.45}
\plotone{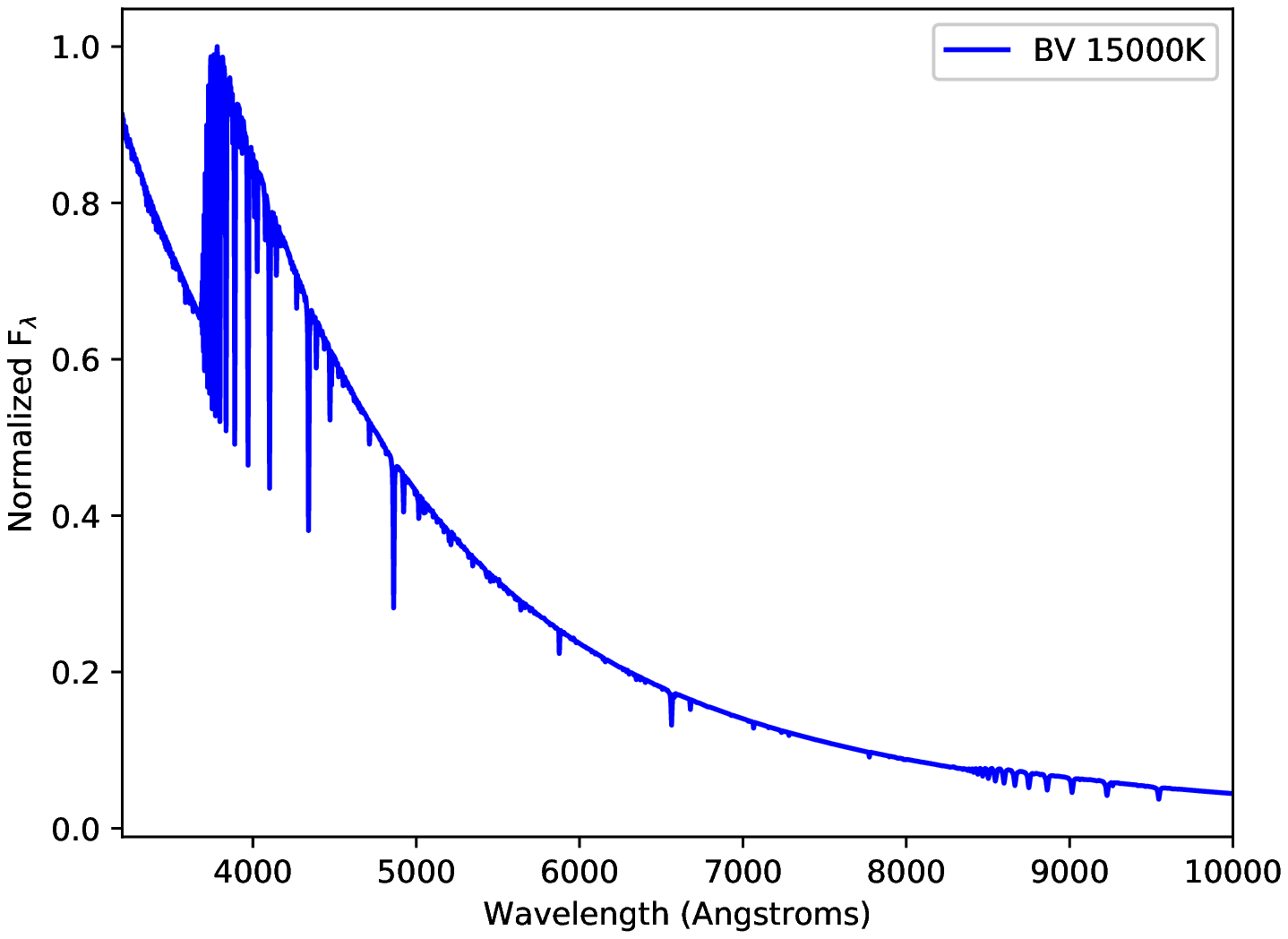}
\plotone{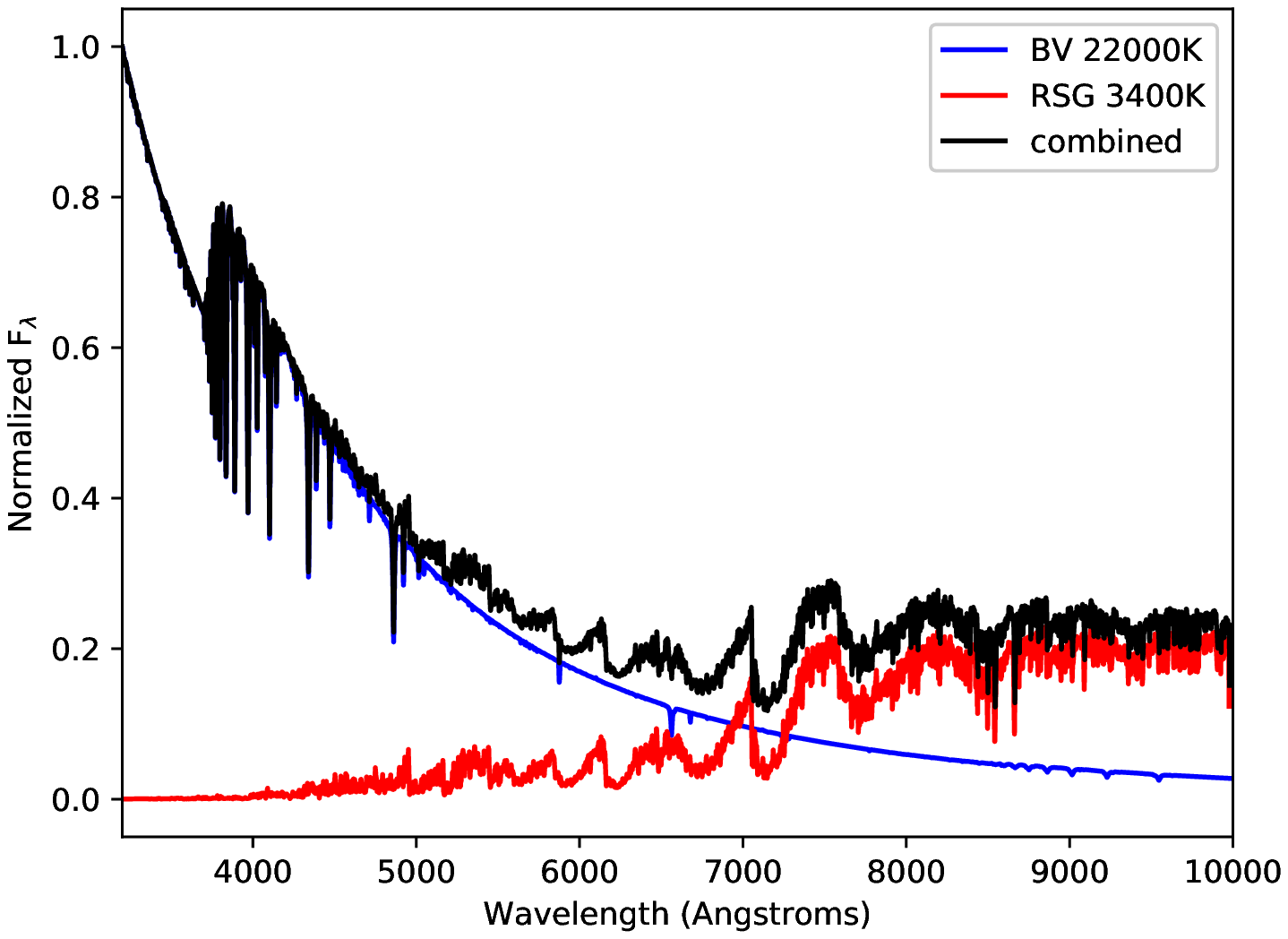}
\plotone{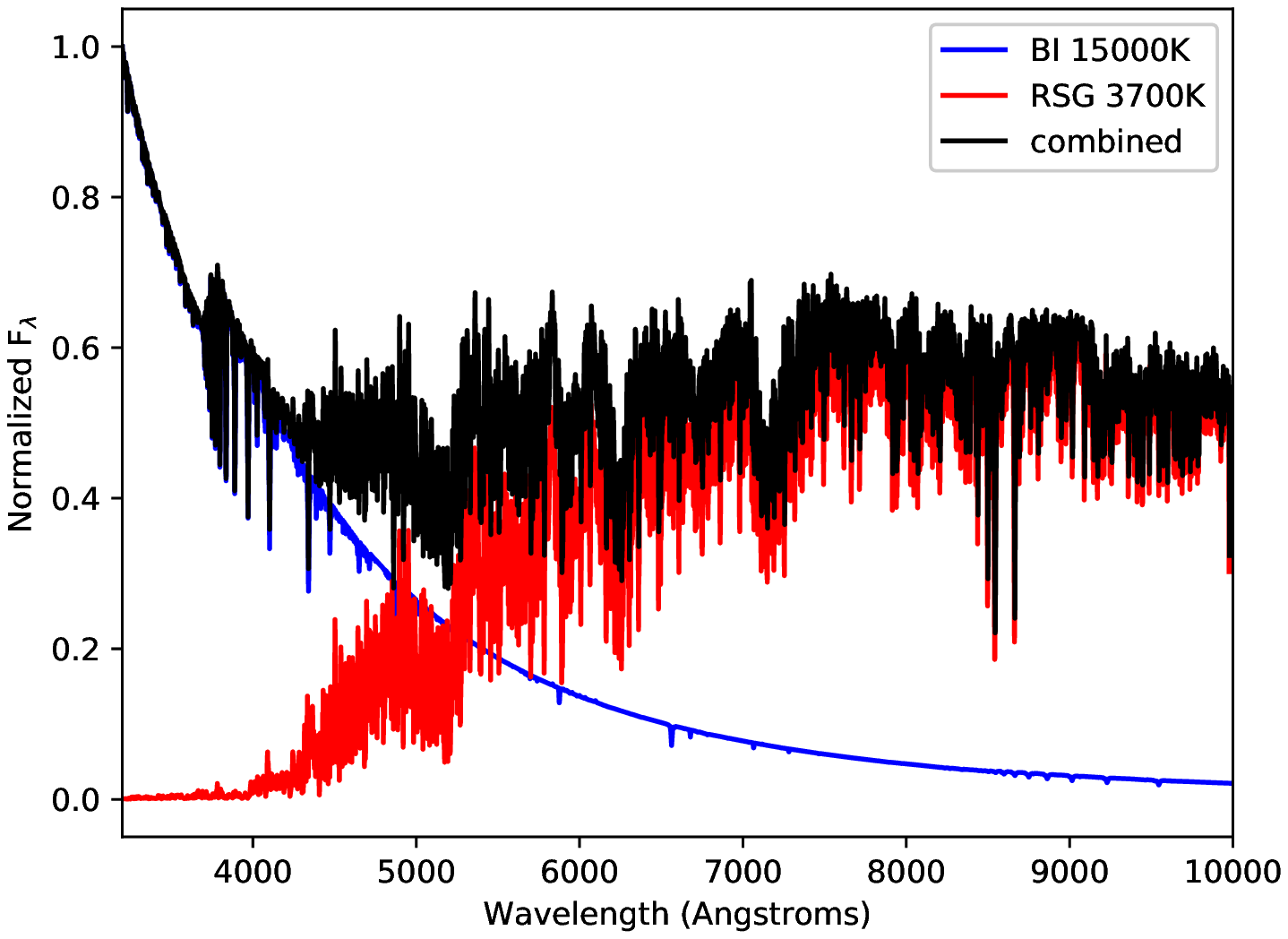}
\plotone{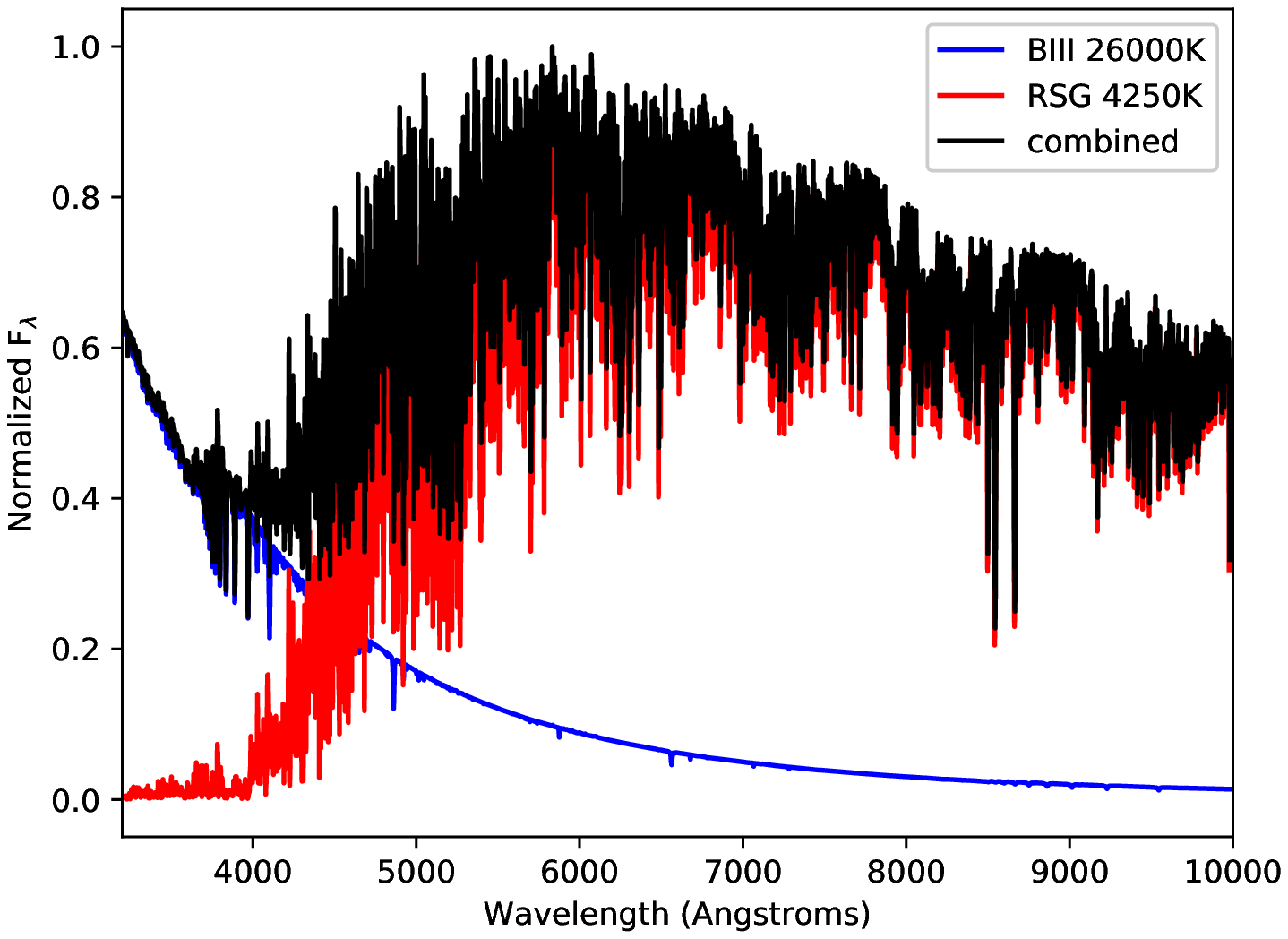}
\plotone{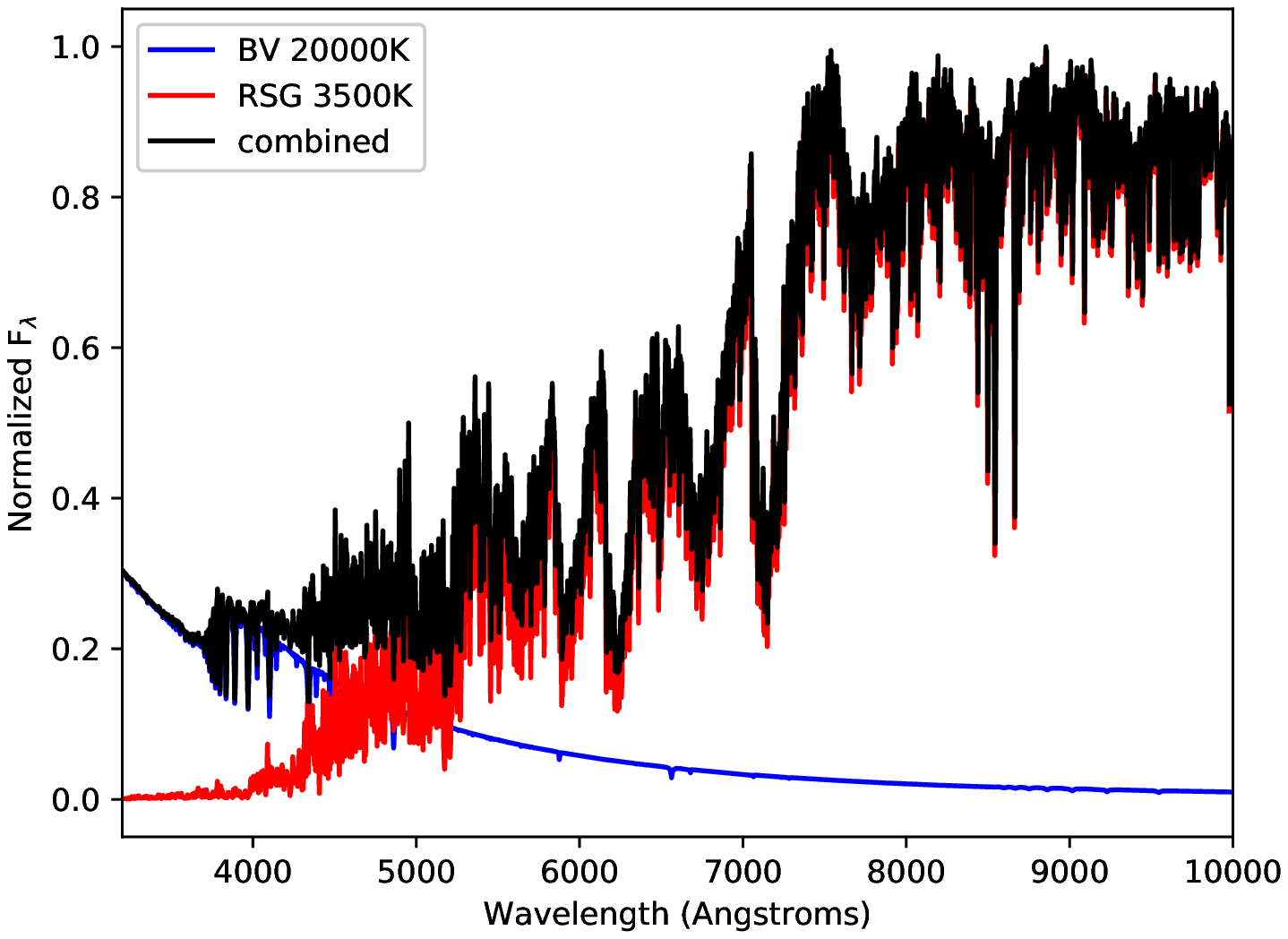}
\plotone{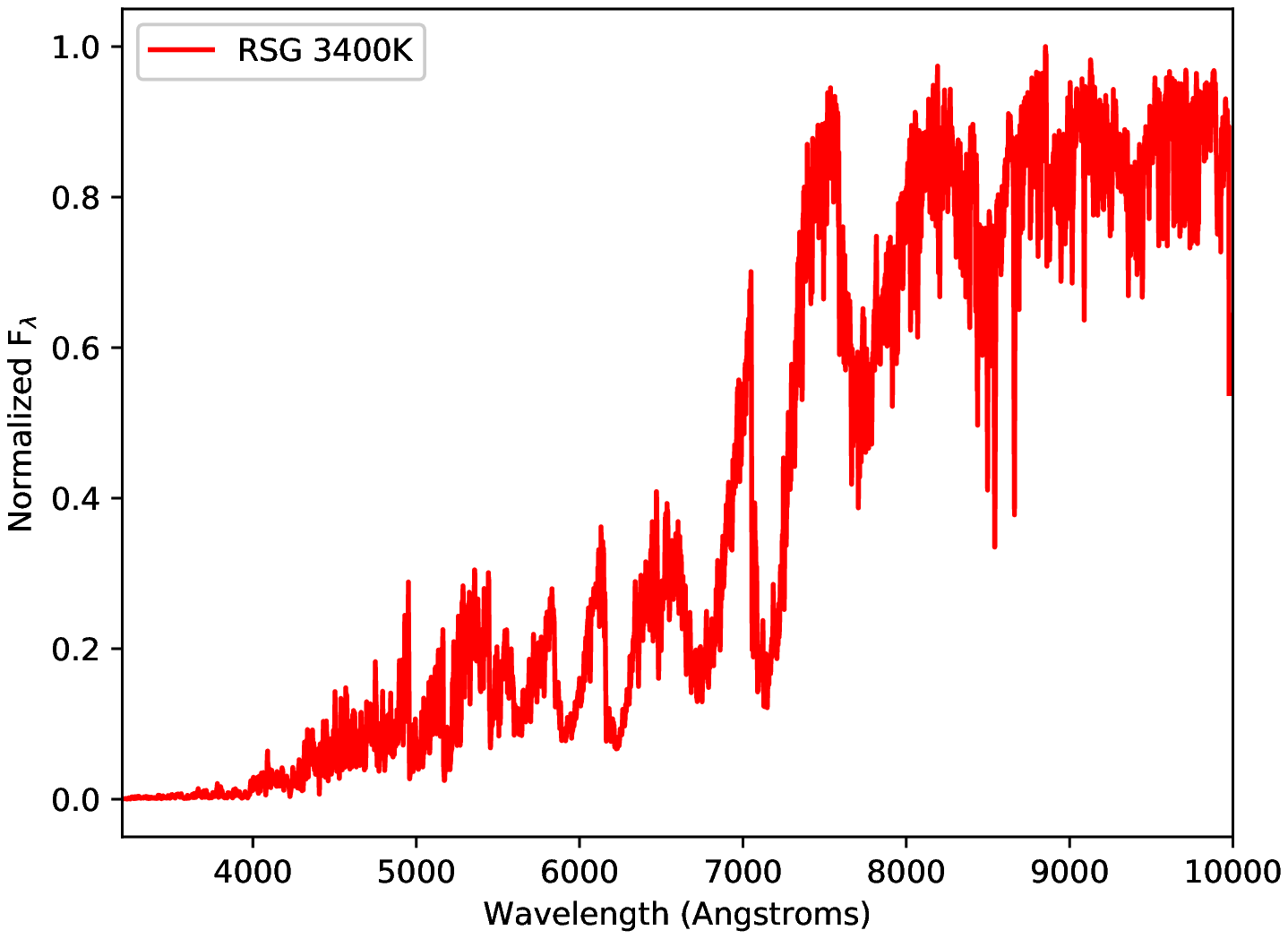}
\caption{\label{fig:SyntheticSpectra} Various combinations of un-reddened synthetic BSTARS06 and MARCS spectra going from a 15000 K BI to a 3400 K RSG with four combinations in between. Note as the Balmer lines grow fainter as the TiO bands grow stronger from top to bottom.}
\end{figure}

The resulting $\sim 25,000$ models span a range of combined spectra. A few examples can be seen in Figure~\ref{fig:SyntheticSpectra}. The model in the top left shows a single 15000 K B dwarf with clear upper Balmer lines. The model in the top right then shows a cool RSG added into a warm B dwarf. The resulting model shows strong Balmer lines but a hint of TiO bands. As the models progress, the TiO bands become stronger and the Balmer lines become weaker as the RSG begins to dominate the spectrum. Finally, the bottom right panel shows a single 3400 K RSG spectrum. 

Now that we have our combined spectra, we can compare our results with observations. 31 Cyg is an  eclipsing binary made up of a K4Iab + B4IV-V with a period of around 10 years \citep{Samus09}. To test the accuracy of the synthetic combined spectra, we created a combined spectrum of a K4I+B4V star (where the K4I is modeled by a 3800 K RSG and the B4V is modeled by a 19000 K BV), and then compared it to the observed spectrum of this system kindly provided by Phil Bennett. Note the similar shape of the spectral energy distribution (SED) as well as the strengths of the upper Balmer lines and TiO feature near 5167\AA.

\section{Observations}
\begin{figure}
\epsscale{1}
\plotone{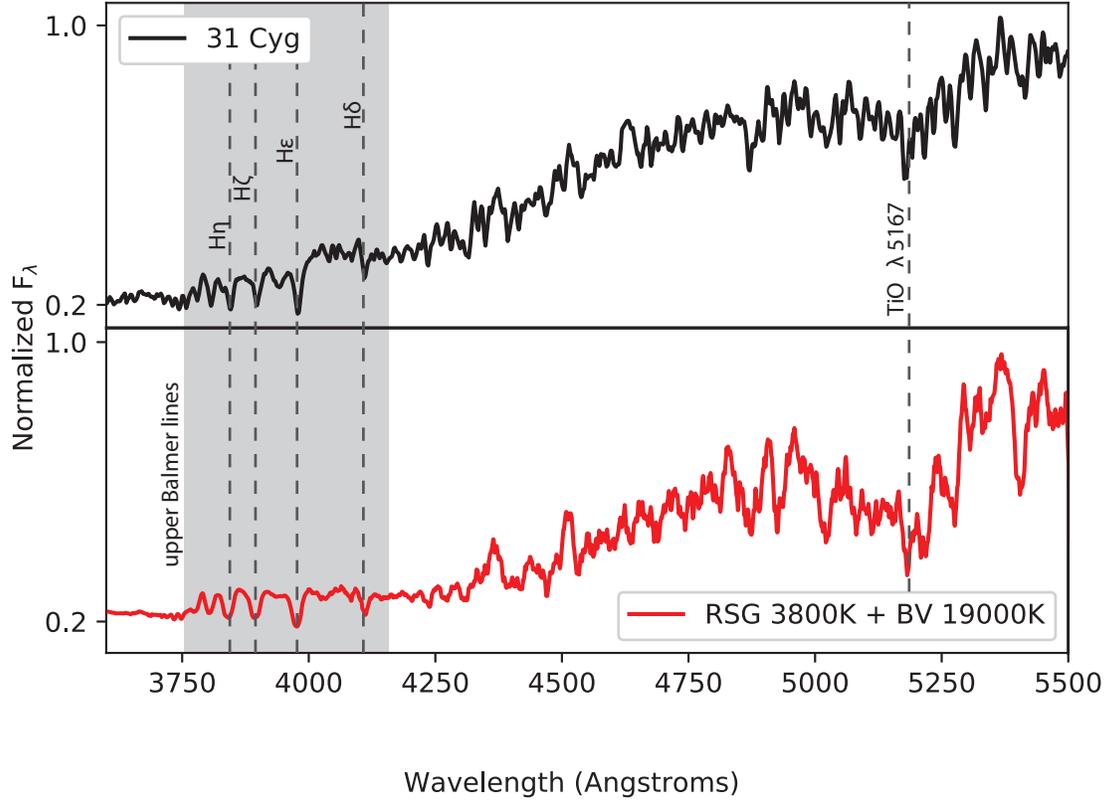}
\caption{\label{fig:31Cyg} Comparison of observed 31 Cyg spectrum with model BSTAR06 + MARCS combined spectrum. 31 Cyg is a K4Iab + B4IV-V so we chose a 3800 K RSG model to represent the K4Iab star and a 19000 K B dwarf model to represent the B4IV-V star. Note the agreement between the upper Balmer lines between 3750 and 4000 \AA\ as well as the strong TiO band at around 5167\AA. Note that the model has been smoothed so that both spectra have about the same resolution. The 31 Cyg spectrum was kindly provided by Phil Bennett.}
\end{figure}

Our synthetic spectra, the spectrum of 31 Cyg (see Figure~\ref{fig:31Cyg}), and the RSG+B stars discussed in the Introduction show that these stars are detectable spectroscopically. From a previous observational survey we had spectra of 598 RSGs in the Magellanic Clouds (MCs) that provided a perfect chance to search for binary RSGs. We also used new spectroscopic observations to search for evidence of additional RSG binaries within our own Galaxy. These observations help us learn more about the binary frequency of RSGs and constrain how difficult they are to discover through spectroscopic observations.

\subsection{Magellanic Cloud Observations}
\begin{figure}
\epsscale{0.48}
\plotone{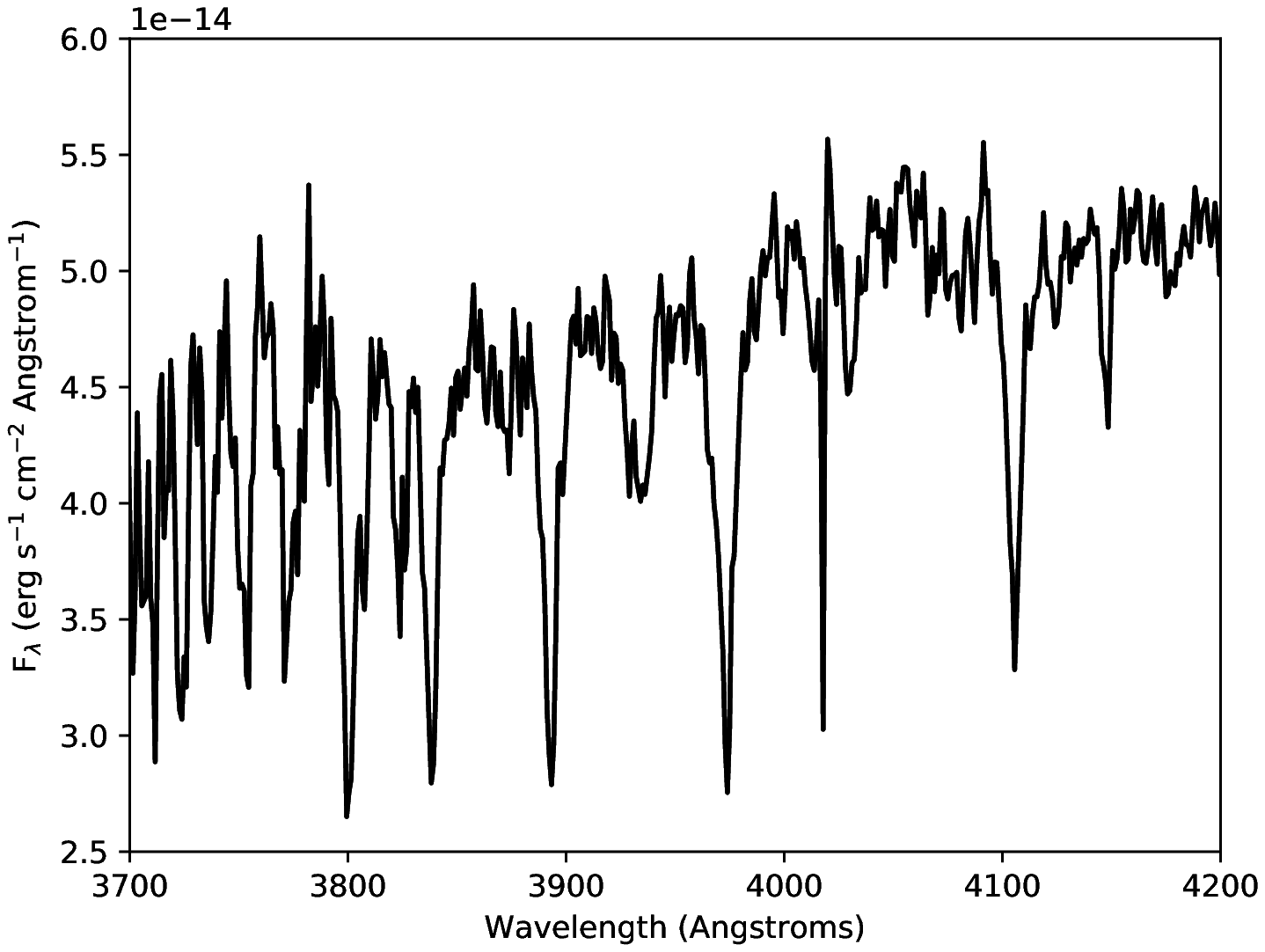}
\plotone{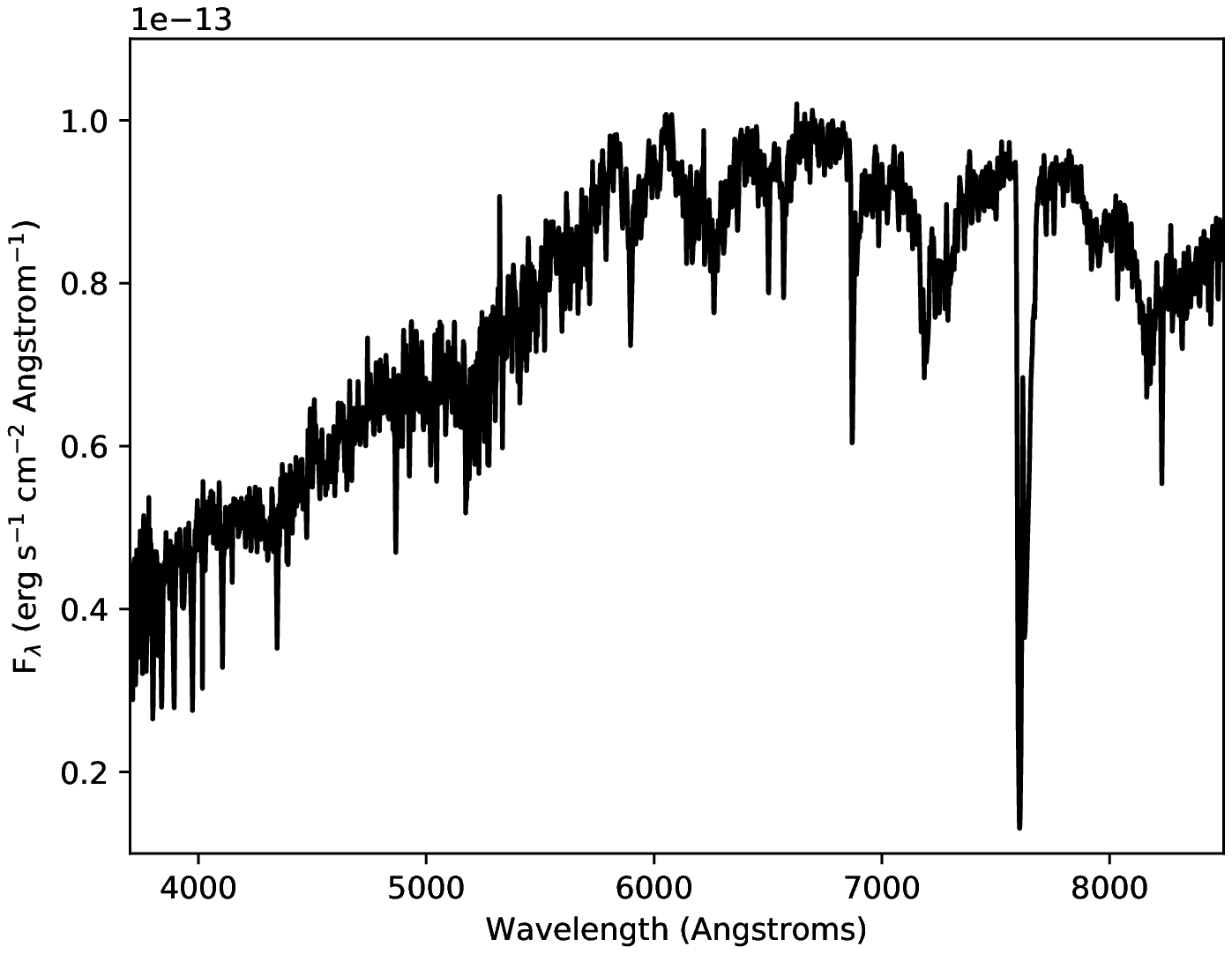}
\caption{\label{fig:KateSpec} Spectrum of the LMC star J05043378-6806235, a 3900 K K1I + B star companion. In the figure on the right, notice the strong TiO lines upward of 5000 \AA\ indicative of a RSG with strong Balmer lines between 3700 and 4000 \AA. The figure on the left shows a closer look at the Balmer lines. Spectra will be published in an upcoming paper by Evans et al.}
\end{figure}

In connection with another project, we obtained spectra of 598 RSGs within the MCs, and classified
them using their spectral lines and determined MC membership based on radial velocities \citep{EvansAAS}. (The spectra will be described in more detail in a future paper by Evans et al.)  We then examined their spectra for this work in an attempt to detect upper Balmer lines. An example spectrum with observable Balmer lines is shown in Figure~\ref{fig:KateSpec}. Of the 311 Small Magellanic Cloud (SMC) stars and 287 Large Magellanic Cloud (LMC) stars, 25 stars (14 LMC and 11 SMC) have clear upper Balmer lines in their spectra as is shown in Figure~\ref{fig:KateSpec}. We did not find any spectra that appeared to be in a spectroscopic binary with any other type of star (O star, RSG, or WR, for example). While these small numbers might not be enough to do population statistics (the LMC and SMC RSG samples cannot be considered complete), it is clear that RSG + B star binaries are present and observable throughout the MCs.

\subsection{Galactic Observations}
We additionally carried out our own observing campaign searching for Balmer lines in the spectra of known Galactic RSGs. Our goal was not to characterize all of the known RSGs within the Galaxy. Instead we wanted to determine how difficult it is to detect different types and luminosity classes of B stars within a RSG's spectrum.

\subsubsection{Observations and Reductions}
Targets were selected from \citet{Humphreys78} and \citet{Levesque05} to ensure we were observing confirmed RSGs as opposed to red giants. Observations were taken on the Apache Point Observatory's 3.5-m telescope using the Dual Imaging Spectrograph (DIS). All of our observations were taken over a clear three night run from UT 21-23 June 2018 during the B half of the night. The seeing was fairly steady and ranged between 1-1.5" throughout the three nights. We observed the 0.9" slit on the first night but switched to the 1.2" slit for the second and third nights due to poor seeing conditions. Since we were interested in the upper Balmer lines spanning from around 3750 - 4000\AA, we used the high dispersion B1200 grating centered at 4280\AA\, giving a total wavelength coverage from 3660\AA\ to 4900\AA at $R=3500$. Because the U-band fluxes of the RSGs are quite faint despite the generally bright $V$ magnitudes, we wound up having to expose for up to 2 hours for a $V=10$ star to achieve a SN of 100 in the region of the upper Balmer lines. When observing these stars we simply turned off the red channel of the camera to avoid almost certain saturation in the red. We observed the stars in order of decreasing magnitude (and thus increasing exposure time) in an attempt to observe the largest number of targets. We additionally observed Feige 110 and BD+28 421 for flux calibration. 

Biases, quartz projector flats, and He-Ne-Ar comparison arcs were obtained during the day and used for reductions. The data were first overscan-subtracted and trimmed and then the remaining bias structure was removed using a master averaged bias. The spectra were extracted using IRAF's ``doslit" package \footnote{IRAF is distributed by the National Optical Astronomy Observatory, which is operated by the Associ- ation of Universities for Research in Astronomy (AURA) under a cooperative agreement with the National Science Foundation.}. Spectra with multiple observations were then combined with cosmic rays rejected. 

\subsubsection{Hidden B Stars}
\begin{figure}
\epsscale{0.5}
\plotone{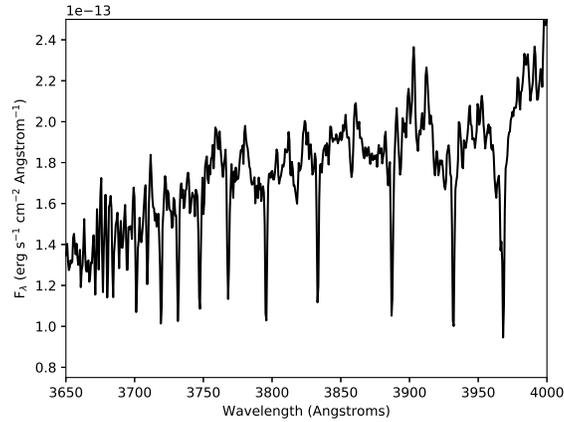}
\caption{\label{fig:VVCep} Observed Balmer lines in W Cep. Notice the strong Balmer lines observed as part of our Galactic RSG observing campaign. This proves to us that our set-up was capable of detecting the bright Balmer lines of a RSG+B star binary.}
\end{figure}

During our observing run we observed 20 known RSGs with S/N $\sim$ 100. None of these stars had Balmer lines except for one known binary, W Cep which is a V = 7.6 mag binary. Our spectrum of W Cep can be seen in Figure~\ref{fig:VVCep} showing that at least one of the brighter known RSG + B stars can be detected through the presence of Balmer absorption lines.

\begin{figure}
\epsscale{0.45}
\plotone{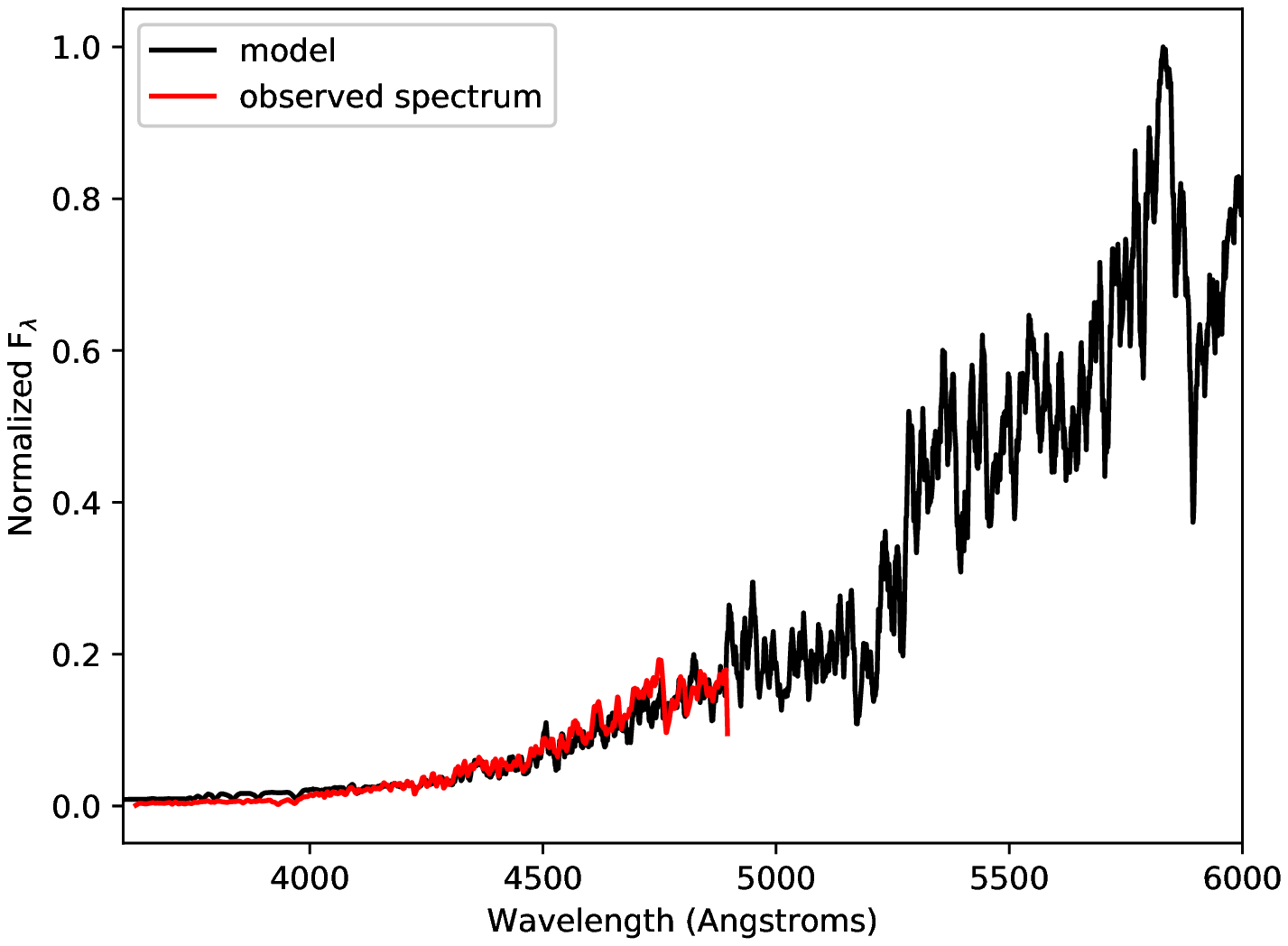}
\plotone{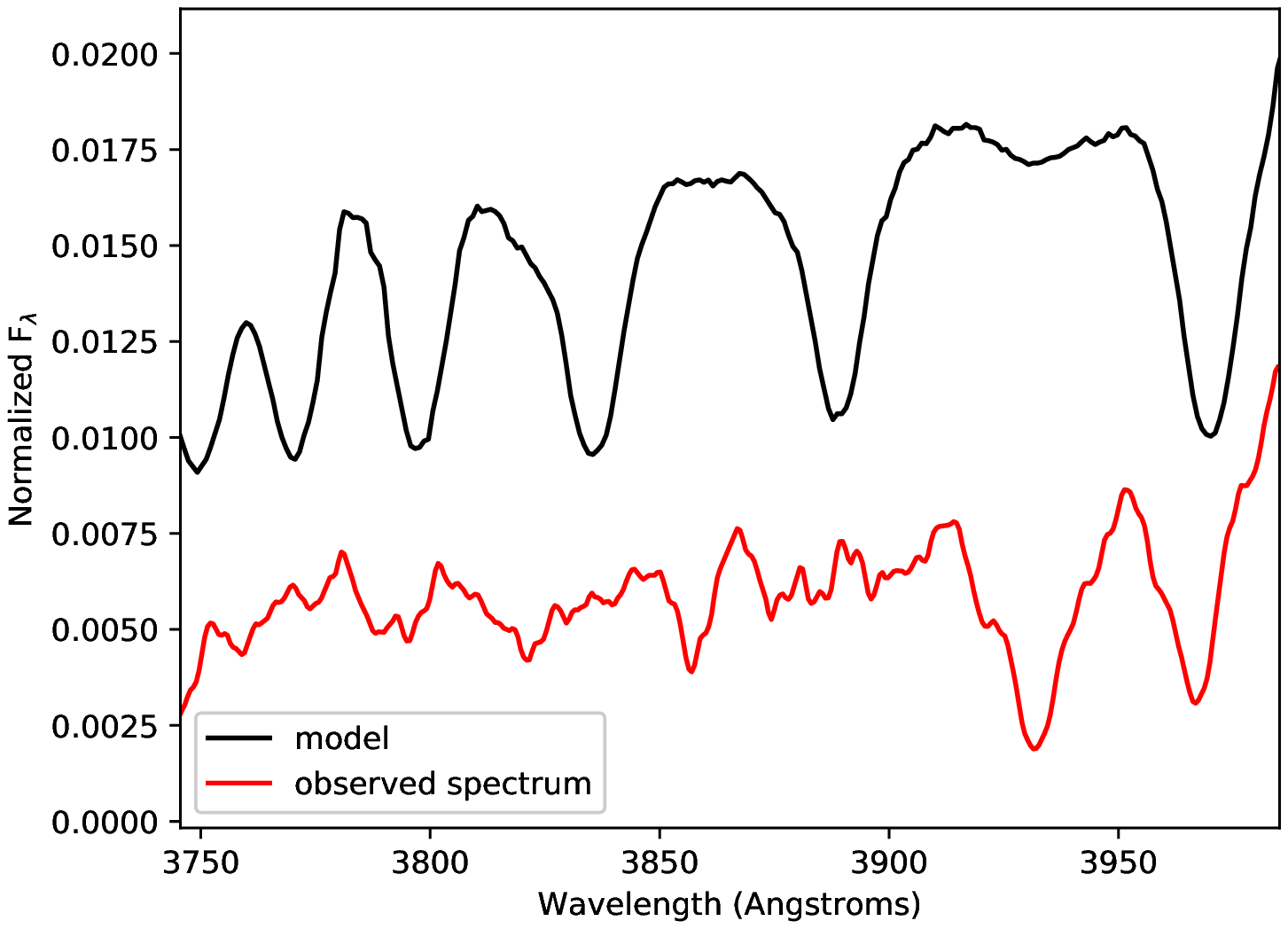}
\caption{\label{fig:hidden} Observed spectrum of BD+39 4208 a 3600 K M3I compared with a cool B dwarf model with Balmer lines. Balmer lines are not visible in the observed spectrum and by comparing them to a spectrum with a similar S/N that contains Balmer lines it is clear that if the star had a B star companion we would have detected the Balmer lines and thus the companion.}
\end{figure}

Even though we did not find any new RSGs with B star companions, these spectra allowed us to constrain the type of B star that could be hidden in the spectra (if any). To do this we relied on the fact that most of these stars had previously been modeled by \citet{Levesque05} using the MARCS models. Because of this, we were able to shift the observed spectra to the appropriate MARCS model and match the two SEDs. We could then add in the blue BSTAR06 models of increasing brightness until it was possible to see Balmer lines. This process is shown in Figure~\ref{fig:hidden}. Of the 20 observed stars, six stars had not been modeled using MARCS spectra. Three of the warmest RSGs could have concealed B dwarf stars up to 30000 K ($M_V = -3.5$), while 11 could not have concealed even the faintest BSTAR06 model in our sample (a 15000 K B dwarf; $M_V = -1.5$). Thus, given a S/N of greater than 100 at moderation spectral resolutions ($R>3000$), it is possible to determine conclusively whether a RSG has a visible (i.e., non-eclipsed) B star companion contributing to its spectrum.

There is a potential problem with this approach. We are making the initial assumption that the star modeled by the MARCS model is a single RSG (with no B star companion) and thus the signal on the blue end is only coming from the RSG. If there was a B star hidden within the spectrum, the blue end would be raised slightly. However, given that we do not see any Balmer lines in the spectrum we believe we can rule out this possibility.

\section{Photometric Criteria for Detecting Companions}
Next we examined the photometric properties of the RSG+B star binaries. Figures~\ref{fig:JHCspec} and \ref{fig:SyntheticSpectra} show the synthetic spectra of a binary RSG and B star. Note that the SED of a RSG+B star binary spectrum is quite a bit flatter than that of either a B star or RSG alone; there is nearly equal signal in both the red and the blue. Given the right set of photometric criteria, we should be able to detect this feature.

\subsection{Synthetic Photometry from Models}
\begin{figure}
\epsscale{0.55}
\plotone{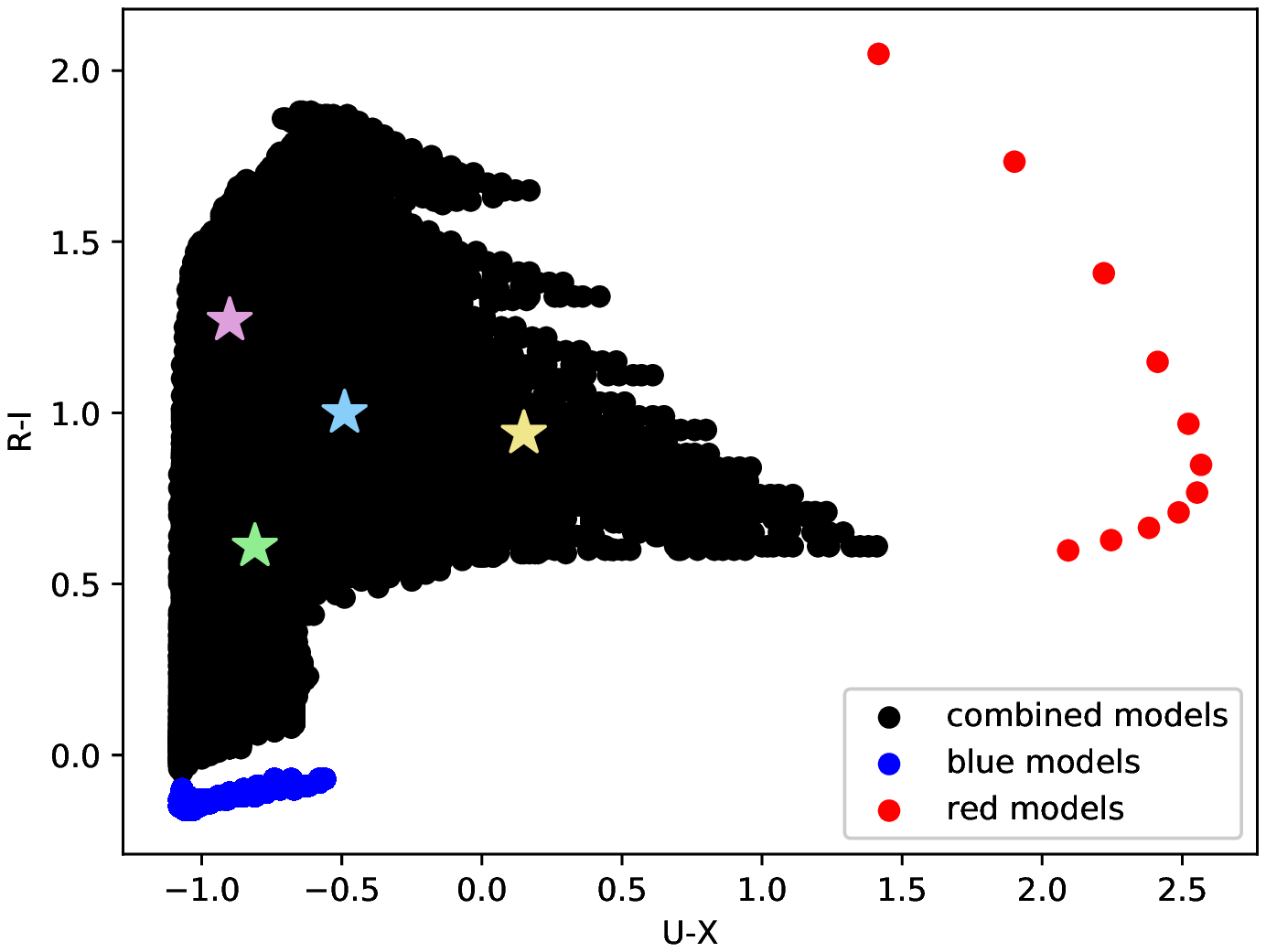}
\plotone{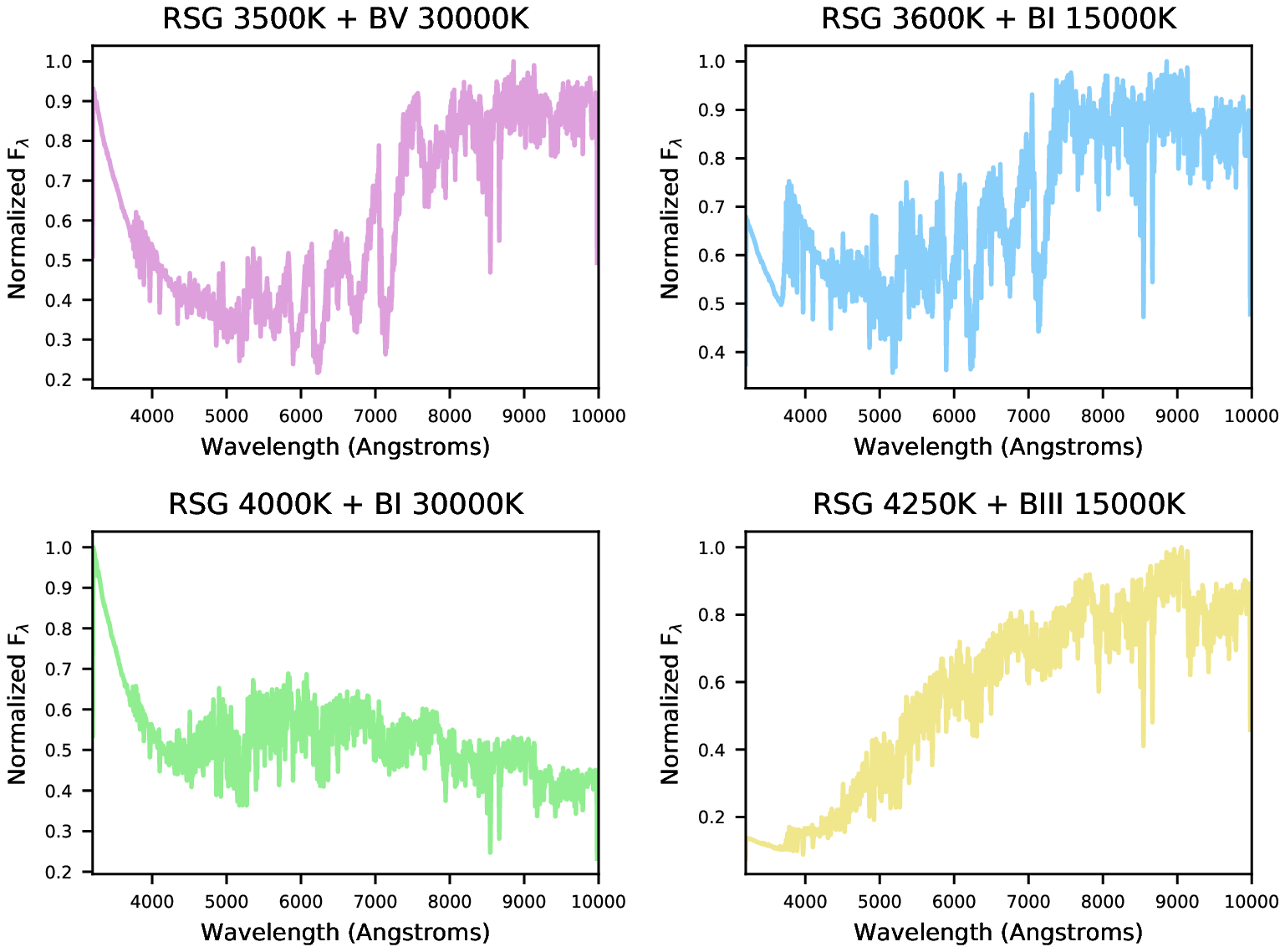}
\caption{\label{fig:SynPhot} Synthetic Photometry of 25,000 models. Blue points represent the B-type star models. Red points represent the RSG models. Black points represent the photometry of the combined synthetic spectra. The pastel-colored stars show example spectra from different regions within the color-color space. Notice that all combined spectra show examples of the Balmer lines as well as RSG-like spectral components. Depending on where the example falls within the composite spectra region, the Balmer lines will be more or less visible. For example, spectra near the red models (yellow star) show more red features than the spectra near the blue models (green star).}
\end{figure}

To investigate this we used our set of $\sim$25,000 binary model spectra and synthetic photometry. To simulate photometry of our spectral models we used the Johnson-Cousins filter curves \citep{Johnson53, Cousins76} along with the zero-points from \citet{Bessell98} and bandpasses from \citet{Bessell90}. We calculated U,B,V,R and I photometry. To best represent the combination of blue and red spectra, we plotted our bluest colors (U-B) versus our reddest colors (R-I) in Figure~\ref{fig:SynPhot}. The figure on the left shows synthetic photometry from single blue stars, single red stars, and the combined spectra. Note that the composite spectra (black) all fall in a single region in this color-color space. The pastel-colored stars on the diagram and plots below show example combined spectra that occupy different regions within color-color space. They all show evidence of a B star's Balmer lines as well as a RSG's TiO bands, with varying strengths depending on where they fall within the black ``composite spectra" region. Spectra near the red models (yellow star on plot) unsurprisingly show more red features than the spectra near the blue models (green star) which show more blue features. 

\subsection{Comparisons with Observed Spectra}
We next wanted to see how well our synthetic photometry matched with the photometry of our observed MC RSG sample discussed in Section 4.1. While we could determine colors based on the spectroscopy, we were limited by the poor quality of the flux calibrations. The MC data were taken using the Anglo-Australian Telescope's multi-fiber fed spectrograph, 2dF+AAOmega. Flux calibration of fiber instruments is notoriously difficult due to differences in the sensitivity function from one fiber to another \citep{Fabricant2008, MasseyHanson}. The fluxing of AAOmega spectra was based on an observation of one spectrophotometric standard taken through a single fiber on a different night than the science data. While the spectral energy distribution looks reasonable for some of the data, it is clearly wrong for others. Thus, we were unsure of any colors derived from spectroscopy. In addition, the spectra did not extend as far in the blue as all of the $U$ band, nor as far into the red as all $I$ band.  Thus, we were unwilling to rely on colors derived from the spectroscopic data.

\begin{figure}
\epsscale{1}
\plotone{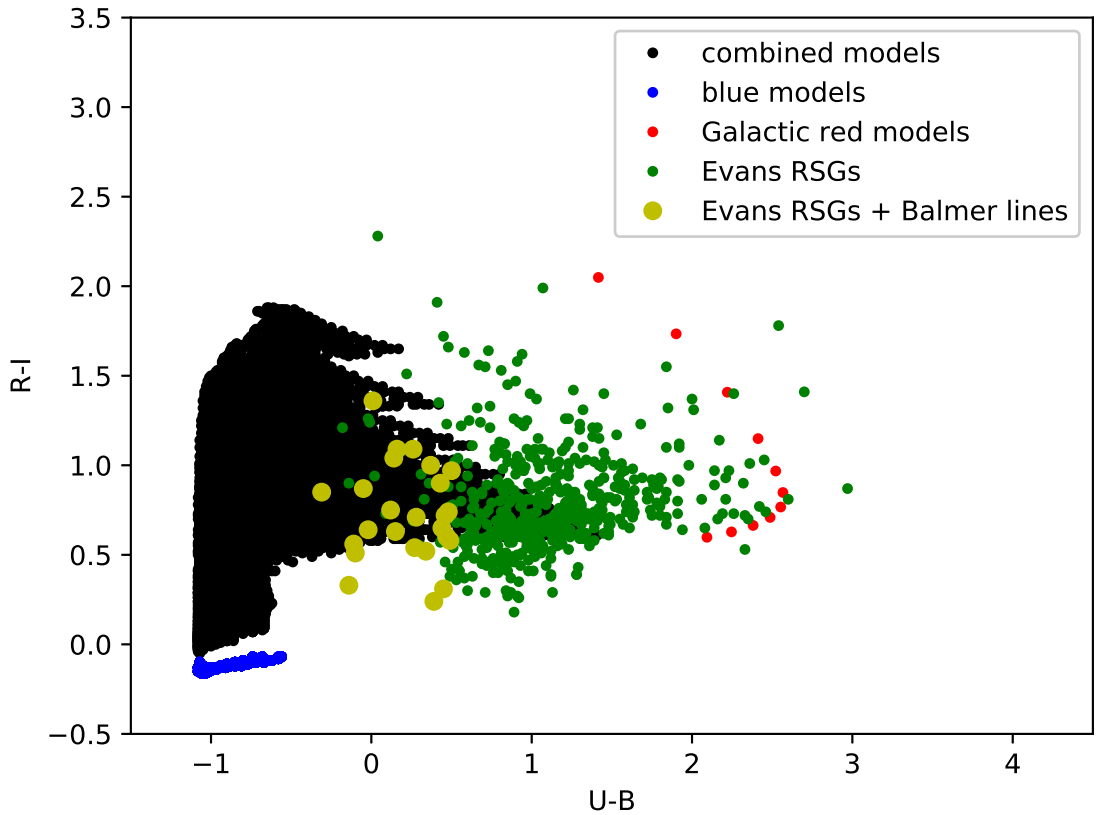}
\caption{\label{fig:KateStars} Evans et al.\ (in prep) stars plotted with the synthetic photometry. Of the $\sim$600 stars observed by Evans et al.\ (in prep), 25 of them (yellow dots) showed evidence of Balmer lines while the rest (green dots) did not. For the most part, these yellow dots fall within the composite spectra region of the photometry from the combined synthetic spectra. Note that there is some contamination of blue and yellow points within the composite spectra region. This is discussed in the text.}
\end{figure}

Figure~\ref{fig:KateStars} shows the result of plotting U-B vs.\ R-I of the observed MC RSGs on top of the BSTAR06 and MARCS model's synthetic photometry. Only the stars with U, B and R magnitudes from \citet{Massey02} and the I magnitudes from \citet{ZaritskySMC} (SMC) or \citet{ZaritskyLMC} (LMC) have been plotted. As discussed earlier, of the 311 SMC stars and 287 LMC stars, 25 stars (14 LMC and 11 SMC) had clear upper Balmer lines in their spectra. In the figure, the 25 stars with observed Balmer lines are within the composite spectra region (with the exception of a few that fall below the region). There is some contamination with non-Balmer-lined RSGs but this contamination is minimal. Most of these spectra show an up-turn in the blue which could be due to either bad photometry, poor flux calibrations, or the physical explanation of possible UV excess \citep{Massey05}, such as with R Coronae Borealis stars \citep{Hecht98}. Hopefully more observations will help explain these (and other) contaminants.

Note that very few of the green points that represent the Magellanic Cloud photometry of single RSGs overlap with the MARCS model's red points. We believe this is due to a failure of the MARCS models to accurately reproduce the observed spectrum in the blue. This was previously found by \citet{Massey05} when looking at the reddening of RSGs. Because the blue MARCS spectrum is somewhat too weak it allows the BSTAR06 models to dominate the blue spectrum earlier than they should. Thus, the upper Balmer lines become visible within the synthetic spectrum before they would in an observed spectrum. This effect does not influence our results negatively. Instead, it simply means that the stars on the left (bluer) side of the composite spectra region are less likely to be binary RSG+B stars than those on the right (redder) side. As will be discussed below, we plan on testing this observationally.

\begin{figure}
\epsscale{1}
\plotone{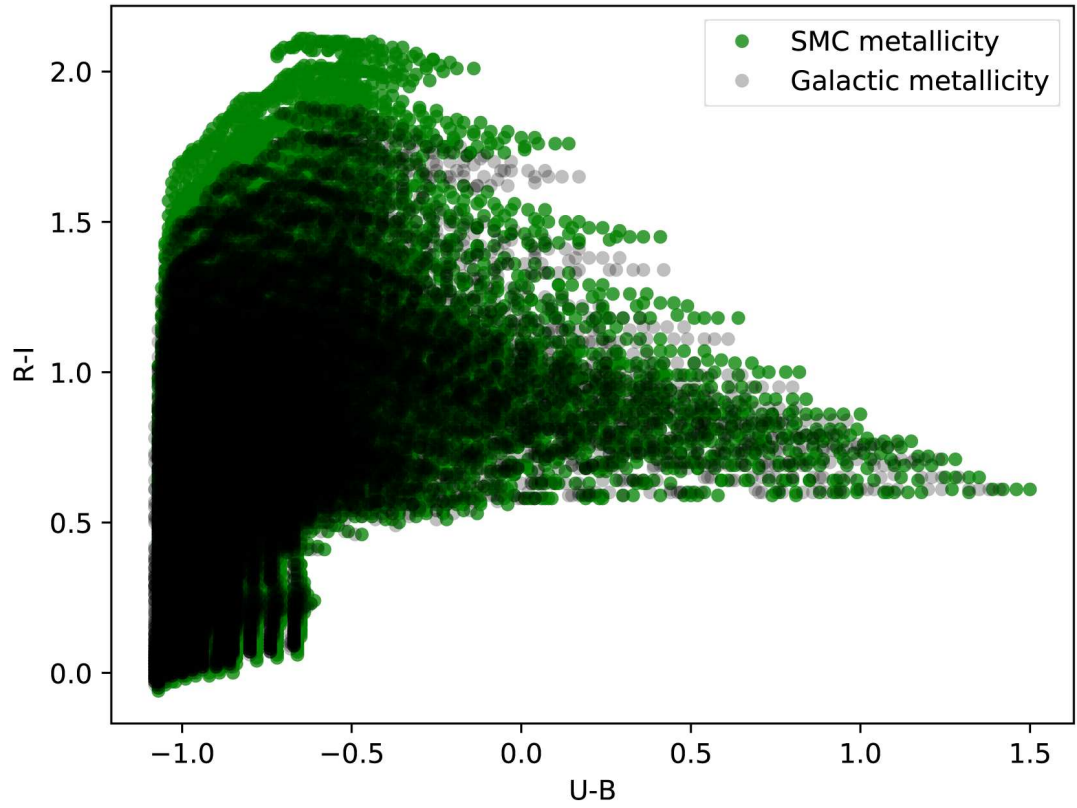}
\caption{\label{fig:metallicity} Synthetic photometry of models with Galactic and SMC metallicities. Changing the metallicity of the synthetic models made only small changes to the overall shape of the combined spectra region in color-color space. Therefore we can use the photometric cutoffs for stars of all metallicities.}
\end{figure}

We were a little concerned about comparing apples to oranges in terms of metallicities. All of the models were computed with solar metallicities ($Z = Z_{\odot}$) while the LMC and SMC have metallicities of $0.5Z_{\odot}$ and $0.27Z_{\odot}$, respectively \citep{RussellDopita}. Figure~\ref{fig:cutoff} shows synthetic photometry from the MARCS RSG models plotted for three different metallicities (solar, LMC, and SMC). As you can see, shifts in the photometry due to metallicity are small. We additionally re-ran the synthetic models using BSTAR06 and the MARCS models with a metallicity of $0.25Z_{\odot}$ (SMC-like) and found the result shown in Figure~\ref{fig:metallicity}. While there is some difference between the synthetic photometry with solar and SMC metallicity, the changes are minor and will not play any role in changing the photometric criteria.

\subsection{Candidate Selection}
\begin{figure}
\epsscale{1}
\plotone{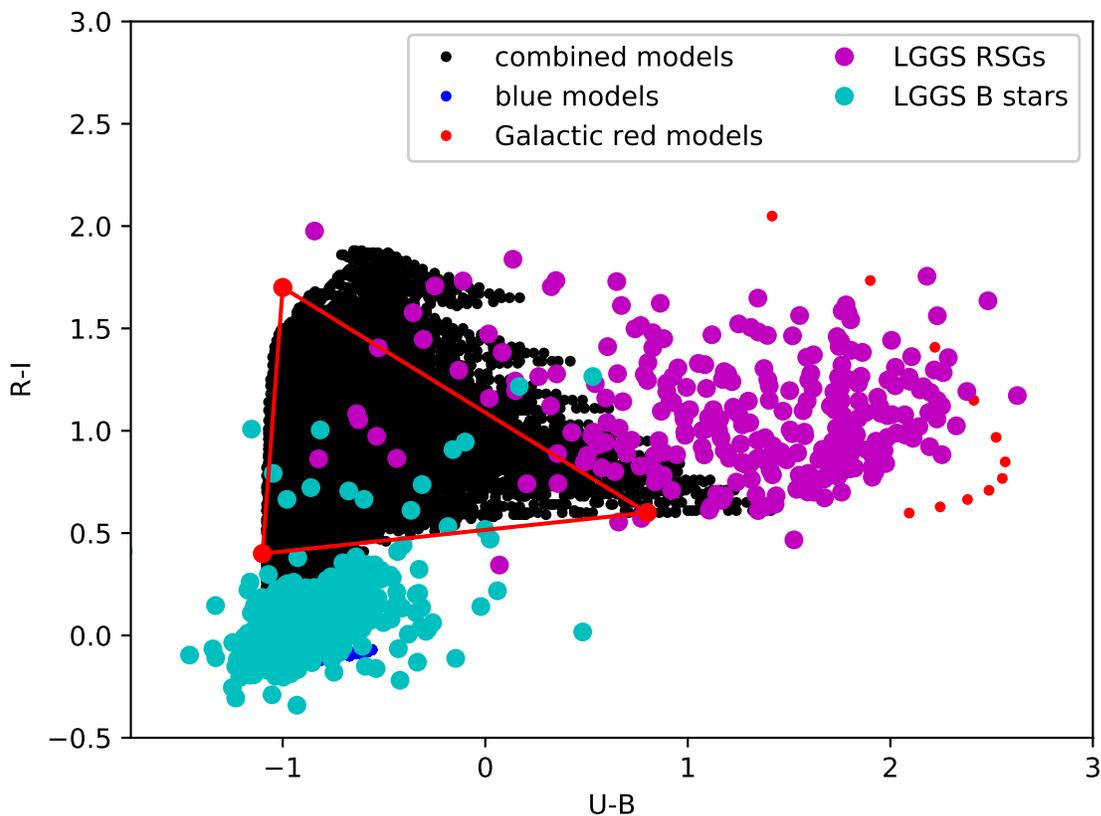}
\caption{\label{fig:LGGSpoints} M31 and M33 LGGS B stars and RSGs plotted with the photometrically selected cutoff region. The blue points are the single B star models and the red points are the single MARCS RSG models. The black points are then the combined models using the Galactic metallicity RSGs. The cyan points are the M31 and M33 LGGS B stars while the magenta points are the M31 and M33 LGGS RSGs. To create as little contamination as possible we have selected the triangle denoted by the red lines as the area encompassing possible RSG+B star binaries in color-color space. The vertices of the points in U-B vs.\ R-I color space are (-1.1,0.4), (0.8,0.6), and (-1,1.7).}
\end{figure}

Given these results we were able to define a list of U-B and R-I photometric cutoffs where we believe RSGs with B star companions should fall on a color-color plot. This region is marked in red in Figure~\ref{fig:LGGSpoints}. B stars and RSGs in M31 and M33 as confirmed by the Local Group Galaxy Survey \citep{LGGS} (LGGS) are plotted as well. The triangle-shaped photometric cutoffs were selected to remove as many B and RSG stars as possible, though there is still a small amount of contamination. 

\begin{figure}
\epsscale{1}
\plotone{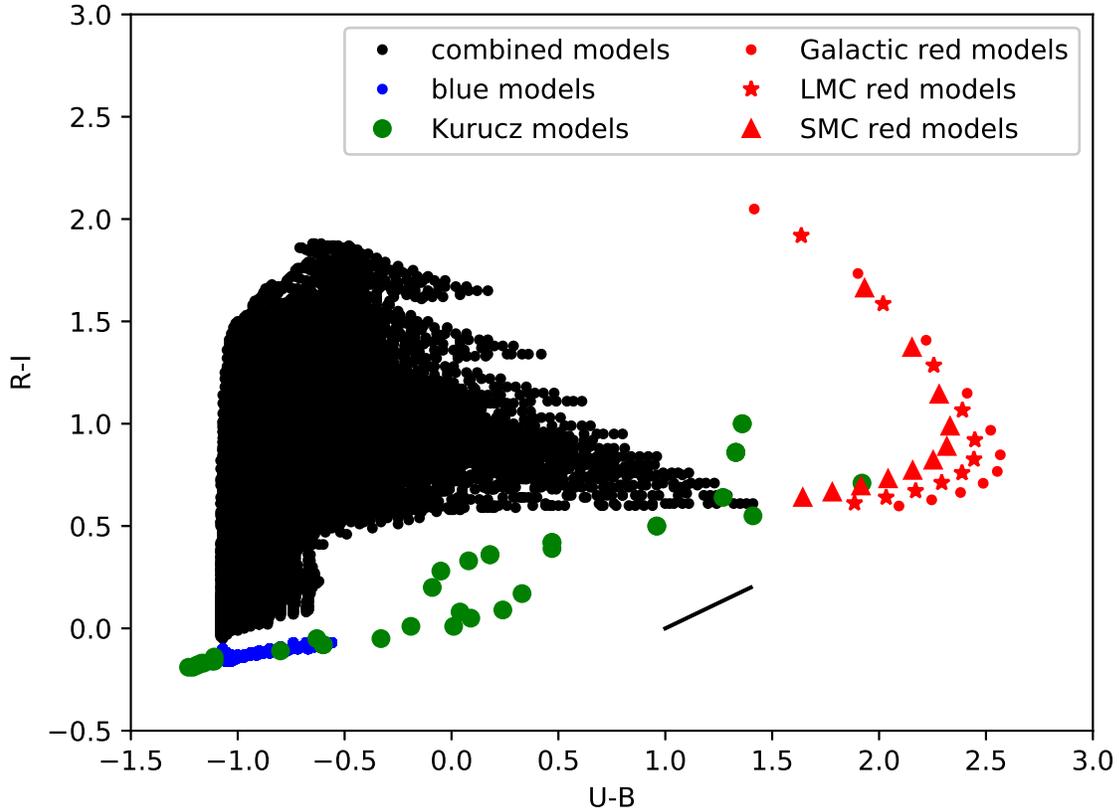}
\caption{\label{fig:cutoff} Effects of metallicity and contamination. The blue points are the single B star models and the red points are the single MARCS RSG models (shown for SMC, LMC and Galactic metallicities). The black points are then the combined models using the Galactic metallicity RSGs. We have additionally plotted the Kurucz models (green circles) for a variety of spectral types showing that we will not have contamination from other types of stars. Finally, we have plotted the reddening vector for $A_V=1$ as a black line showing that the reddened Kurucz models will not enter the region of composite spectra.}
\end{figure}

We also investigated contamination from other spectral types and reddening. Figure~\ref{fig:cutoff} shows photometry from the Kurucz models \citep{Kurucz}. These model atmospheres span all spectral types with a large range in effective temperatures and surface gravities. We chose a model that represented each spectral type and luminosity class and plotted their colors on the U-B vs.\ R-I plot. As is shown in the figure, none of these spectral types fall within the selected region (or within the composite spectra region in general) and thus contamination by stars with other spectral types should not be an issue. Reddening should also not be a problem. Figure~\ref{fig:cutoff} shows an $A_V=1$ reddening vector. This simply pushes all of the stars slightly up and to the right. Given this reddening vector, none of the Kurucz models will enter the photometrically selected region.

Given these parameters we were able to build a candidate list of potential RSG+B star binaries for stars within the LGGS, specifically within M31 and M33. To do this, we filtered through the 370,000 stars in M31 and 150,000 stars in M33. We were first concerned about contamination by red dwarfs. We started with a magnitude cutoff of V=20 to limit the search to the brighter stars (more likely to be RSGs). We then used a color-color diagram to pick out the red dwarfs from the RSGs using U-B vs.\ U-R color cuts. Since we want to eventually get spectra of these stars, we next restricted ourselves to only the isolated stars (as marked in the LGGS). Finally, we weeded out foreground stars by looking for potential matches with \textit{Gaia} \citep{Gaia} and restricted the stars to distances less than 10kpc. At this point we were left with 138 potential binaries in M31 and 142 in M33\footnote{A more detailed list of these stars will be included in an upcoming paper by Neugent et al.\ once follow-up observations have been made.}.

We additionally checked these stars for signal within the UV and NIR. For this we relied on the Panchromatic Hubble Andromeda Treasury (PHAT) dataset \citep{PHAT}. As opposed to sorting through the photometric catalogs, we instead downloaded the image files and searched for the UV (F275W) and NIR (F160W) signatures visually. In all 280 cases, these stars were visually bright in both the NIR and UV suggesting a SED much like what we have predicted. Observations are currently underway, and will be discussed in a subsequent paper.

\section{Conclusions and Future Work}
Using the color criteria described by \citet{MasseyM31RSGs} and \cite{DroutM33RSGs} for separating likely foreground red stars from {\it bona-fide} RSGs, there are 1837 stars brighter than $V\leq20$ whose photometry in the LGGS is consistent with their being RSGs in M31 and M33. This can be compared to the 280 RSG+B star binary candidates we identified as part of this survey. This 15\% value is lower than the observed binary fraction of other massive stars which has been placed at a minimum of 30\%. However, until we have followed up our sample selection with spectroscopy (our next step), we will be unable to accurately estimate our completeness and contamination. 

There are additional scenarios that might increase the binary frequency of RSGs that have not been accounted for as part of this survey. For example a RSG+O star binary is possible from an evolutionary point of view. However, based on the Geneva evolutionary models, it is not as common as a RSG+B star. Equal mass RSG+RSG binaries should also exist and in rarer cases WR+RSG binary should form when the RSG star has an initial lower mass (note that we do not expect binary systems with stars less massive than B stars since an A-type dwarf will not yet have arrived on the main sequence before the RSG has ended its life.) There are additionally RSG+B star binaries whose signatures are difficult to detect because the lower-mass companion has effectively been consumed by the RSG and is completely concealed, at least in terms of direct spectroscopic or photometric signatures of its photosphere. An example of this star is VV Cep whose binarity is only visible when the stars are out of eclipse. These stars will not be detectable using the method we have proposed in this paper.

Ultimately we hope to gain a better grasp of the binary fraction of RSGs. To do this we first need to determine how many binary RSG systems are detectable using different methods. Given the huge overhead of radial velocity searches, spectroscopy of a complete sample of RSGs, and period determination through light curves, we have come up with a method of answering this question photometrically that will hopefully cut down on the resources needed to detect binary RSGs, and make it possible to search for these systems in future large-scale or automated photometric surveys.

\acknowledgements
We would like to thank both Jamie Lomax and the incredibly helpful Apache Point Observatory support staff that assisted in training K.\ Neugent on the APO 3.5m. Additionally, conversations with Phil Bennett about various aspects of RSG binaries proved to be very useful. Comments from the anonymous referee further improved this paper. This work was supported in part by NSF IGERT grant DGE-1258485 as well as by a fellowship from the Alfred P.\ Sloan Foundation. P.M.'s work was supported by the National Science Foundation under grant AST-1612874.

	
\bibliographystyle{apj}
\bibliography{masterbib}

\end{document}